\documentclass[twocolumn,preprintnumbers,amsmath,amssymb]{revtex4}

\usepackage{graphicx}
\usepackage{dcolumn}
\usepackage{bm}

\begin{document}

\newcommand{\fourqquads}{{\qquad \qquad \qquad \qquad}}
\newcommand{\eightqquads}{{\qquad \qquad \qquad \qquad \qquad \qquad \qquad \qquad}}
\newcommand{\fournegspc}{{\! \! \! \!}}
\newcommand{\eightnegspc}{{\! \! \! \! \! \! \! \!}}
\newcommand{\AGAt}{${\rm Al_{0.3}Ga_{0.7}As}$}
\newcommand{\AGAff}{${\rm Al_{0.55}Ga_{0.45}As}$}
\newcommand{\AGAy}{${\rm Al_{\it y}Ga_{1-{\it y}}As}$}
\newcommand{\AGAx}{${\rm Al_{\it x}Ga_{1-{\it x}}As}$}
\newcommand{\AGAf}{${\rm Al_{0.15}Ga_{0.85}As}$}
\newcommand{\IGAx}{${\rm In_{0.53}Ga_{0.47}As/In_{0.52}Al_{0.48}As}$}
\newcommand{\IGx}{${\rm In_{0.53}Ga_{0.47}As}$}
\newcommand{\AGA}{GaAs/AlGaAs}
\newcommand{\IGA}{InGaAs/InAlAs}
\newcommand{\WcmK}{W/cm$\cdot$K}
\newcommand{\E}[1]{$E_{#1}$}
\newcommand{\N}[1]{$N_{#1}$}
\newcommand{\DN}{$\Delta N_{32}$}
\newcommand{\taut}[2]{$\tau_{#1}^{#2}$}
\newcommand{\Schrodinger}{Schr\"{o}dinger}
\newcommand{\epmiwt}{{e^{\pm i \omega t}}}
\newcommand{\empiwt}{{e^{\mp i \omega t}}}
\newcommand{\eiwt}{{e^{i \omega t}}}
\newcommand{\epiwt}{{e^{+i \omega t}}}
\newcommand{\emiwt}{{e^{-i \omega t}}}
\newcommand{\tauul}{{\tau_{u\rightarrow l}}}
\newcommand{\hw}{{\hbar \omega}}
\newcommand{\hwz}{{\hbar \omega_0}}
\newcommand{\cc}{{\mathrm{c.c.}}}
\newcommand{\Ef}{{\mathcal{E}}}
\newcommand{\Pol}{{\mathcal{P}}}
\newcommand{\Rdiff}{{\mathcal{R}}}
\newcommand{\Rdiffm}{{\mathcal{R}_\mathrm{th}^-}}
\newcommand{\Rdiffp}{{\mathcal{R}_\mathrm{th}^+}}
\newcommand{\dw}{{d \omega}}
\newcommand{\hwab}{{\hbar \omega_{a b}}}
\newcommand{\Np}{{N_\mathrm{p}}}
\newcommand{\Pout}{{P_\mathrm{out}}}
\newcommand{\GammabyNp}{{\frac{\Gamma}{\Np}}}
\newcommand{\etaIth}{{\eta_{I\mathrm{th}}}}
\newcommand{\etaI}{{\eta_I}}
\newcommand{\nst}{{n^\mathrm{st}_\mathrm{ph}}}
\newcommand{\nsp}{{n^\mathrm{sp}_\mathrm{ph}}}
\newcommand{\nph}{{n_\mathrm{ph}}}
\newcommand{\tpara}{{\tau_{3,\mathrm{para}}}}
\newcommand{\tsp}{{\tau_\mathrm{sp}}}
\newcommand{\tph}{{\tau_\mathrm{ph}}}
\newcommand{\tpsp}{{\tau'_\mathrm{sp}}}
\newcommand{\tst}{{\tau_\mathrm{st}}}
\newcommand{\gmat}{{g_{\mathrm{mat}}}}
\newcommand{\Ith}{{I_{\mathrm{th}}}}
\newcommand{\Ithp}{{I_{\mathrm{th}}^+}}
\newcommand{\Ithm}{{I_{\mathrm{th}}^-}}
\newcommand{\Jth}{{J_{\mathrm{th}}}}
\newcommand{\Jthf}{{J_{\mathrm{th},5 \mathrm{K}}}}
\newcommand{\Jmaxf}{{J_{\mathrm{max},5 \mathrm{K}}}}
\newcommand{\Tactive}{{T_{\mathrm{active}}}}
\newcommand{\Tcw}{{T_{\mathrm{cw}}}}
\newcommand{\Tmax}{{T_{\mathrm{max}}}}
\newcommand{\Tmaxp}{{T_{\mathrm{max,pul}}}}
\newcommand{\Tmaxc}{{T_{\mathrm{max,cw}}}}
\newcommand{\Jmax}{{J_{\mathrm{max}}}}
\newcommand{\Jpeak}{{J_{\mathrm{peak}}}}
\newcommand{\Imax}{{I_{\mathrm{max}}}}
\newcommand{\Vac}{{V_{\mathrm{ac}}}}
\newcommand{\Vmod}{{V_{\mathrm{mod}}}}
\newcommand{\Aac}{{A_{\mathrm{ac}}}}
\newcommand{\Vcav}{{V_{\mathrm{cav}}}}
\newcommand{\Acav}{{A_{\mathrm{cav}}}}
\newcommand{\SEQUAL}{\mbox{SEQUAL}}
\newcommand{\mez}{{m^{\ast}(z)}}
\newcommand{\me}{m^\ast}
\newcommand{\e}{m^{\ast}(z,E)}
\newcommand{\mezei}{m^{\ast}(z,E_i)}
\newcommand{\mezef}{m^{\ast}(z,E_f)}
\newcommand{\ddz}{\frac{\partial}{\partial z}}
\newcommand{\pddt}{\frac{\partial}{\partial t}}
\newcommand{\ddt}{\frac{d}{d t}}
\newcommand{\etal}{{\it {et~al.\ }}}
\newcommand{\ie}{\mbox{i.\ e.}}
\newcommand{\psicz}{\psi_c(z)}
\newcommand{\psiciz}{\psi_c^{(i)}(z)}
\newcommand{\psicfz}{\psi_c^{(f)}(z)}
\newcommand{\epsc}{\epsilon_{\mathrm{core}}}
\newcommand{\epsmc}{\epsilon_{\mathrm{m,core}}}
\newcommand{\epsr}{\epsilon_{\mathrm{r}}}
\newcommand{\epsm}{\epsilon_{\mathrm{m}}}
\newcommand{\epsd}{\epsilon_{\mathrm{d}}}
\newcommand{\epsdt}{\epsilon^2_{\mathrm{d}}}
\newcommand{\epsz}{\epsilon_{0}}
\newcommand{\km}{k_\mathrm{m}}
\newcommand{\kd}{k_\mathrm{d}}
\newcommand{\kdt}{k^2_\mathrm{d}}
\newcommand{\degC}[1]{{#1\,^\circ\mathrm{C}}}
\newcommand{\betaz}{{\beta_z}}
\newcommand{\neff}{{n_\mathrm{eff}}}
\newcommand{\wpl}{{\omega_{\mathrm{p}}}}
\newcommand{\wplt}{{\omega^2_{\mathrm{p}}}}
\newcommand{\wspl}{{\omega_{\mathrm{sp}}}}
\newcommand{\kpari}{{\mathbf{k}_{\|,i}}}
\newcommand{\kparf}{{\mathbf{k}_{\|,f}}}
\newcommand{\kparm}{{\mathbf{k}_{\|,m}}}
\newcommand{\kparn}{{\mathbf{k}_{\|,n}}}
\newcommand{\kpara}{{\mathbf{k}_{\|,a}}}
\newcommand{\kparb}{{\mathbf{k}_{\|,b}}}
\newcommand{\ikpari}{{i,{\mathbf k}_{\|,i}}}
\newcommand{\fkparf}{{f,{\mathbf k}_{\|,f}}}
\newcommand{\mkparm}{{m,{\mathbf k}_{\|,m}}}
\newcommand{\akpara}{{a,{\mathbf k}_{\|,a}}}
\newcommand{\bkparb}{{b,{\mathbf k}_{\|,b}}}
\newcommand{\phinr}{{\phi_n \rangle}}
\newcommand{\phiml}{{\langle \phi_m}}
\newcommand{\psitket}{{|\psi(t)\rangle}}
\newcommand{\iket}{{| i \rangle}}
\newcommand{\jket}{{| j \rangle}}
\newcommand{\aket}{{| a \rangle}}
\newcommand{\bket}{{| b \rangle}}
\newcommand{\ket}[1]{{| #1 \rangle}}
\newcommand{\bra}[1]{{\langle #1 |}}
\newcommand{\ibra}{{\langle i |}}
\newcommand{\jbra}{{\langle j |}}
\newcommand{\abra}{{\langle a |}}
\newcommand{\bbra}{{\langle b |}}
\newcommand{\kpar}{{\mathbf{k}_{\|}}}
\newcommand{\ahat}{{\hat{a}}}
\newcommand{\Vbias}{{V_\mathrm{bias}}}
\newcommand{\ahatc}{{\hat{a}^\dagger}}
\newcommand{\eikr}{{e^{i \mathbf{k}\cdot\mathbf{r}}}}
\newcommand{\emikr}{{e^{-i \mathbf{k}\cdot\mathbf{r}}}}
\newcommand{\br}{{\mathbf{r}}}
\newcommand{\bz}{{\mathbf{z}}}
\newcommand{\bk}{{\mathbf{k}}}
\newcommand{\bu}{{\mathbf{u}}}
\newcommand{\usigma}{{\mathbf{u}_\sigma}}
\newcommand{\ksigma}{{\mathbf{k},\sigma}}
\newcommand{\rpar}{{\mathbf{r}_\|}}
\newcommand{\bA}{{\mathbf A}}
\newcommand{\bE}{{\mathbf E}}
\newcommand{\ki}{{\mathbf{k}_i}}
\newcommand{\kf}{{\mathbf{k}_f}}
\newcommand{\kj}{{\mathbf{k}_j}}
\newcommand{\kg}{{\mathbf{k}_g}}
\newcommand{\qpara}{{\bf q}_{\parallel}}
\newcommand{\qz}{{\bf q}_z}
\newcommand{\LO}{{\mathrm{LO}}}
\newcommand{\hwLO}{{\hbar\omega_{\mathrm{LO}}}}
\newcommand{\wLO}{{\omega_{\mathrm{LO}}}}
\newcommand{\ELO}{{E_{\mathrm{LO}}}}
\newcommand{\bracomboa}{{| n_a^\ksigma : \akpara \rangle}}
\newcommand{\bracombob}{{| n_b^\ksigma : \bkparb \rangle}}
\newcommand{\comboa}{{n_a^\ksigma : \akpara}}
\newcommand{\combob}{{n_b^\ksigma : \bkparb}}
\newcommand{\mat}{{\mathrm{mat}}}
\newcommand{\rmmin}{{\mathrm{min}}}
\newcommand{\rmmax}{{\mathrm{max}}}
\newcommand{\rmpeak}{{\mathrm{peak}}}
\newcommand{\rmph}{{\mathrm{ph}}}
\newcommand{\rmps}{{\mathrm{ps}}}
\newcommand{\rmst}{{\mathrm{st}}}
\newcommand{\rmsp}{{\mathrm{sp}}}
\newcommand{\thz}{{\mathrm{THz}}}
\newcommand{\meV}{{\mathrm{meV}}}
\newcommand{\eV}{{\mathrm{eV}}}
\newcommand{\ps}{{\mathrm{ps}}}
\newcommand{\rmand}{{\mathrm{and}}}
\newcommand{\rmwhere}{{\mathrm{where}}}
\newcommand{\rmpara}{{\mathrm{para}}}
\newcommand{\cav}{{\mathrm{cav}}}
\newcommand{\thD}{{\mathrm{3D}}}
\newcommand{\twD}{{\mathrm{2D}}}
\newcommand{\rmeff}{{\mathrm{eff}}}
\newcommand{\rmhot}{{\mathrm{hot}}}
\newcommand{\rmLO}{{\mathrm{LO}}}
\newcommand{\rmmW}{{\mathrm{mW}}}
\newcommand{\rmuW}{{\mu\mathrm{W}}}
\newcommand{\rmK}{{\mathrm{K}}}
\newcommand{\nGaAs}{{n_\mathrm{GaAs}}}
\newcommand{\nSi}{{n_\mathrm{Si}}}
\newcommand{\rmGaAs}{{\mathrm{GaAs}}}
\newcommand{\slopeeff}{{\frac{\alpha_{\rmm,\rmf}}{\alpha_\rmw+\alpha_{\rmm,\rmf}+\alpha_{\rmm,\rmr}}}}
\newcommand{\rmfor}{{\mathrm{for}}}
\newcommand{\inlimit}{{\mathrm{in \ the \ limit}}}
\newcommand{\rmtb}{{\mathrm{tb}}}
\newcommand{\rmr}{{\mathrm{r}}}
\newcommand{\rmf}{{\mathrm{f}}}
\newcommand{\rme}{{\mathrm{e}}}
\newcommand{\rmR}{{\mathrm{R}}}
\newcommand{\rmi}{{\mathrm{i}}}
\newcommand{\rmnm}{{\mathrm{nm}}}
\newcommand{\rmu}{{\mathrm{u}}}
\newcommand{\rml}{{\mathrm{l}}}
\newcommand{\rmj}{{\mathrm{j}}}
\newcommand{\rmm}{{\mathrm{m}}}
\newcommand{\rmw}{{\mathrm{w}}}
\newcommand{\rmmm}{{\mathrm{mm}}}
\newcommand{\rmmat}{{\mathrm{mat}}}
\newcommand{\rmth}{{\mathrm{th}}}
\newcommand{\rmV}{{\mathrm{V}}}
\newcommand{\defas}{{\stackrel{\triangle}{=}}}
\newcommand{\itof}{{i \rightarrow f}}
\newcommand{\ftoi}{{f \rightarrow i}}
\newcommand{\atob}{{a \rightarrow b}}
\newcommand{\btoa}{{b \rightarrow a}}
\newcommand{\qsig}{{{\bf q},\sigma}}
\newcommand{\epol}{{\bf \hat{e}}_{{\bf q},\sigma}}
\newcommand{\zhat}{{\bf \hat{z}}}
\newcommand{\xhat}{{\bf \hat{x}}}
\newcommand{\pop}{{\hat{\mathbf{p}}}}
\newcommand{\rop}{{\hat{\mathbf{r}}}}
\newcommand{\rzop}{{\hat{\mathbf{z}}}}
\newcommand{\rpop}{{\hat{\mathbf{r}}_{\|}}}
\newcommand{\Aop}{{\hat{\mathbf{A}}}}
\newcommand{\Eop}{{\hat{\mathbf{E}}}}
\newcommand{\Hop}{{\hat{H}}}
\newcommand{\Oop}{{\hat{O}}}
\newcommand{\Omat}{{\bar{\bar{O}}}}
\newcommand{\Htbop}{{\hat{H}_\mathrm{tb}}}
\newcommand{\Htbmat}{{\bar{\bar{H}}_\mathrm{tb}}}
\newcommand{\Hext}{{\hat{H}}}
\newcommand{\Hextop}{{\hat{H}_\mathrm{ext}}}
\newcommand{\Hextmat}{{\bar{\bar{H}}_\mathrm{ext}}}
\newcommand{\Hmat}{{\bar{\bar{H}}}}
\newcommand{\rhomat}{{\bar{\bar{\rho}}}}
\newcommand{\rhoop}{{\hat{\rho}}}
\newcommand{\ntot}{{n_\mathrm{tot}}}
\newcommand{\Lmod}{{L_\mathrm{mod}}}
\newcommand{\Lcav}{{L_\mathrm{cav}}}
\newcommand{\tulhotLO}{{\tau^\rmhot_{ul,\rmLO}}}
\newcommand{\tulLO}{{\tau_{ul,\rmLO}}}
\newcommand{\tFLparb}{{\tau_{5\rightarrow(6,3,2,1)}}}
\newcommand{\tFLpar}{{\tau_{5,\mathrm{par}}}}
\newcommand{\Tpure}{{T^*_2}}
\newcommand{\td}{{\tau_{\|}}}
\newcommand{\tdgen}[1]{{\tau_{\|#1}}}
\newcommand{\tdtw}{{\tau_{\|,2}}}
\newcommand{\tdtwp}{{\tau_{\|,2'}}}
\newcommand{\tdtwpsq}{{\tau^2_{\|,2'}}}
\newcommand{\tdtwth}{{\tau_{\|,23}}}
\newcommand{\tdtwpth}{{\tau_{\|,2'3}}}
\newcommand{\tdth}{{\tau_{\|,3}}}
\newcommand{\tdsq}{{\tau^2_{\|}}}
\newcommand{\Omsqgen}[1]{{\Omega_{#1}^2}}
\newcommand{\Delsqgen}[1]{{\Delta_{#1}^2}}
\newcommand{\tdsqgen}[1]{{\tau^2_{\|#1}}}
\newcommand{\Ahsqgen}[1]{{\left(\frac{\Delta_{#1}}{\hbar}\right)^2}}
\newcommand{\Ahsq}{{\left(\frac{\Delta_0}{\hbar}\right)^2}}
\newcommand{\Ahcsq}{{\left(\frac{\delzc}{\hbar}\right)^2}}
\newcommand{\Ehsqgen}[1]{{\left(\frac{E_{#1}}{\hbar}\right)^2}}
\newcommand{\Ehsq}{{\left(\frac{E_{1'3}}{\hbar}\right)^2}}
\newcommand{\Ehcsq}{{\left(\frac{E_{22'}}{\hbar}\right)^2}}
\newcommand{\delop}{{\hat{\nabla}}}
\newcommand{\zzif}{|z_{i \rightarrow f}|^2}
\newcommand{\deltakr}{{\delta^{\mathrm{kr}}}}
\newcommand{\delnu}{{\Delta \nu}}
\newcommand{\delfwhm}{{\Delta \nu_\mathrm{FWHM}}}
\newcommand{\Erad}{{E_\mathrm{rad}}}
\newcommand{\frad}{{f_\mathrm{rad}}}
\newcommand{\zrad}{{z_\mathrm{rad}}}
\newcommand{\nurad}{{\nu_\mathrm{rad}}}
\newcommand{\delec}{{\Delta E_\mathrm{c}}}
\newcommand{\delnth}{{\Delta n_\mathrm{th}}}
\newcommand{\drr}{{\Delta \mathcal{R}_\mathrm{th}/\mathcal{R}_\mathrm{th}}}
\newcommand{\drrfrac}{{\frac{\Delta \mathcal{R}_\mathrm{th}}{\mathcal{R}_\mathrm{th}}}}
\newcommand{\Rth}{{\mathcal{R}_\mathrm{th}}}
\newcommand{\delzc}{{\Delta^{\mathrm{c}}_0}}
\newcommand{\gth}{{g_\mathrm{th}}}
\newcommand{\delN}{{\Delta N}}
\newcommand{\icm}{{{\mathrm{cm}}^{-1}}}
\newcommand{\Appcm}{{{\mathrm{A}/\mathrm{cm}}^{2}}}
\newcommand{\iicm}{{{\mathrm{cm}}^{-2}}}
\newcommand{\iiicm}{{{\mathrm{cm}}^{-3}}}
\newcommand{\iiium}{{{\mu\mathrm{m}}^{-3}}}
\newcommand{\iiim}{{{\mathrm{m}}^{-3}}}
\newcommand{\kacmm}{{\rm kA/cm}^2}
\newcommand{\real}[1]{{\mathcal{R}\mathrm{e}\{#1\}}}
\newcommand{\imag}[1]{{\mathcal{I}\mathrm{m}\{#1\}}}
\newcommand{\um}{{\mu\mathrm{m}}}
\newcommand{\IV}{{$I$-$V$}}
\newcommand{\VIs}{{$V$-$I$s}}
\newcommand{\IVs}{{$I$-$V$s}}
\newcommand{\VI}{{$V$-$I$}}
\newcommand{\LI}{{$L$-$I$}}
\newcommand{\LIs}{{$L$-$I$s}}
\newcommand{\LVs}{{$L$-$V$s}}
\newcommand{\PI}{{$P$-$I$}}
\newcommand{\LV}{{$L$-$V$}}
\newcommand{\RV}{{$\mathcal{R}$-$V$}}
\newcommand{\RVs}{{$\mathcal{R}$-$V$s}}
\newcommand{\RI}{{$\mathcal{R}$-$I$}}
\newcommand{\RIs}{{$\mathcal{R}$-$I$s}}
\newcommand{\JV}{{$J$-$V$}}
\newcommand{\GV}{{$G$-$V$}}
\newcommand{\CV}{{$C$-$V$}}
\newcommand{\IB}{{$I$-$B$}}
\newcommand{\GB}{{$G$-$B$}}
\newcommand{\dVdI}{{$dV/dI$}}
\newcommand{\dVdIV}{{$dV/dI$-$V$}}
\newcommand{\dVdII}{{$dV/dI$-$I$}}
\newcommand{\ebar}{\overline{\eta}}
\newcommand{\Perot}{{P\'{e}rot}}
\newcommand{\phos}{{$\mathrm{H_3PO_4:H_2O_2:H_2O}$}}
\newcommand{\sulf}{{$\mathrm{H_2SO_4:H_2O_2:H_2O}$}}
\newcommand{\amm}{{$\mathrm{NH_4OH:H_2O_2:H_2O}$}}
\newcommand{\ammlap}{{$\mathrm{NH_4OH:H_2O_2}$}}
\newcommand{\alumina}{{$\mathrm{Al_2O_3}$}}
\newcommand{\alcoat}{{$\mathrm{Al_2O_3/Ti/Au/Al_2O_3}$}}
\newcommand{\ang}{{\mathrm{\AA}}}

\preprint{}

\title{Coherence of resonant-tunneling transport in terahertz quantum-cascade lasers\\}

\author{Sushil Kumar}
\author{Qing Hu}%
\affiliation{%
Department of Electrical Engineering and Computer Science and Research Laboratory of Electronics,
Massachusetts Institute of Technology, Cambridge, MA 02139\\
}%

\date{\today}
\begin{abstract}
We develop simple density-matrix models to describe the role of coherence in
resonant-tunneling (RT) transport of quantum-cascade lasers (QCLs). Specifically,
we investigate the effects of coherent coupling between the lasing levels with
other levels on the transport properties and gain spectra. In the first part
of the paper, we use a three-level density-matrix model to obtain useful analytical
expressions for current transport through the injector barrier in a QCL. An expression
for the slope discontinuity in the current-voltage characteristics at the lasing
threshold is derived. This value is shown to be a direct measure of the population
inversion at threshold, and contradicts the previously held belief of it being indicative
of ratio of the laser level lifetimes. In the second part of the paper, we use
density matrices to compute the gain spectrum for a resonant-phonon terahertz QCL design.
The large anticrossing of the doublet of lower radiative levels is reflected
in a broad gain linewidth due to a coherent RT assisted depopulation process. At certain
bias conditions, the gain spectrum exhibits double peaks which is supported by
experimental observations.

\end{abstract}

\maketitle

\section{\label{sec:intro}Introduction}

The role of sequential resonant-tunneling (RT) in semiconductor quantum-wells for
charge transport~\cite{esaki:sl} as well as light amplification~\cite{kazarinov:sl1}
has been discussed since the early 1970s. Subsequently, multiple proposals were introduced in
the 1980s to predict intersubband lasing in semiconductor superlattice structures. With the
advent of a unique RT based charge injection scheme, a unipolar mid-infrared (IR) intersubband
laser, the so-called quantum-cascade laser (QCL), was first demonstrated in 1994~\cite{faist:qcl}.
Mid-IR QCLs have since undergone rapid progress and room-temperature continuous-wave operation is
now obtained over the \mbox{$\lambda\sim4-10~\um$} spectrum. Based on similar design principles,
the operating frequency range of these devices has also been extended to terahertz frequencies
($\nu\sim1-5~\thz$, $\lambda\sim60-300~\um$)~\cite{kohler:laser}. However, due to additional
challenges associated with the charge transport in these low-frequency devices, terahertz QCLs
are still required to be cryogenically cooled~\cite{williams:review,kumar:review08,scalari:review}.
It then becomes increasingly important to better understand the transport processes in
terahertz QCLs to improve the existing designs, with a goal to ultimately achieve
 room-temperature operation.

The best temperature performance for terahertz QCLs is obtained with the resonant-phonon (RP)
designs~\cite{williams:laser,hu:respho}, and operation up to $186~\rmK$ without magnetic
field~\cite{kumar:diagonal}, and $225~\rmK$ with magnetic field~\cite{wade:thzqcl} has been
demonstrated. For QCLs, almost universally across all different designs and operating frequencies,
 RT plays an important role in the electrical injection process to populate the upper radiative
level~\cite{sirtori:tunnel,terazzi:tunnel}. For the resonant-phonon terahertz QCLs (RPTQCLs),
RT also plays an important role in the lower level depopulation. Newer designs with RT based
extraction have also been demonstrated recently achieving very low threshold current
densities~\cite{scalari:tunnel,scalari:stepwell}.  
The radiative levels in a terahertz QCL have small energy separation in the
range of $\hbar\omega\sim4-20~\meV$ ($\nu\sim1-5~\thz$). The potential barriers in the multiple
quantum-well (QW) structure are therefore kept thick to obtain tightly-coupled levels in order
to maintain selectivity of the injection as well as the depopulation processes. This makes the
coupling energies (as characterized by the anticrossing energy $2\hbar\Omega_{ij}$ for the coupled levels
$i$ and $j$) similar in value to the low-temperature energy level broadening, which is expected to
be of the order of a few $\meV$. Consequently, the loss of coherence (or dephasing) in the RT process
has a significant bearing on the electron transport across the barriers~\cite{callebaut:mcdm}.    

In this paper, simple density-matrix models are used to incorporate the important role of
coherence in the RT transport of QCLs. We consider the case of tight binding for
intermodule transport, since the intermodule energy anticrossings $2\hbar\Omega_{ij}$ are typically
much smaller than the intramodule energy level separations, where each module can consist
of one or more QWs and the QCL structure is a periodic repetition of one or more such modules.
In the first part of the paper, we use a density-matrix model similar to the 2-level model
first proposed by Kazarinov and Suris~\cite{kazarinov:sl2} and then later also discussed in
Refs.~\onlinecite{sirtori:tunnel,willenberg:bloch,williams:thesis} among others.
We modify the model to extend to 3 levels to describe a QCL in a more general sense, and derive
analytical expressions for current transport across the injector barrier both below and above the
lasing threshold. Despite the simplicity of the model, we can gain useful information about
QCL operation from the derived results that includes information about the possible temperature
degradation mechanisms in RPTQCLs. The results presented in this section
are applicable to mid-IR QCLs as well. In the second part of the paper, we use a
similar model to estimate the optical gain spectrum of RPTQCLs. A coherent RT assisted
depopulation process is stipulated to be the cause of broad gain bandwidths that are
typically observed in these lasers~\cite{kumar:surfemit}.

\begin{figure}[htbp]
\centering
\includegraphics[width=2.2in]{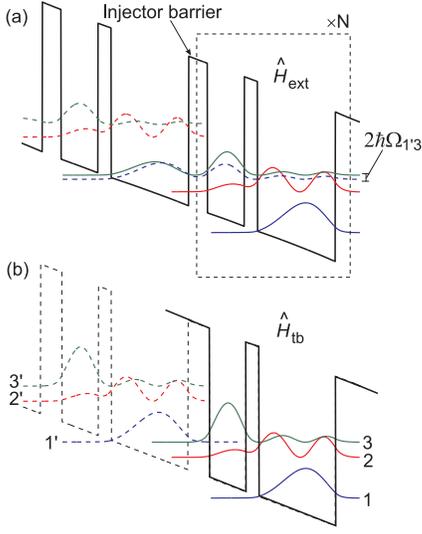}
\caption{
Plot of the magnitude squared envelope wavefunctions for a 3-level QCL design
with two different sets of basis functions. Plot (a) is for the ``extended'' scheme, where
energy splitting due to the injector anticrossing $2\hbar\Omega_{1'3}$ is visible, and the
wavefunctions are calculated for a potential profile $\Hextop$ as it appears in the figure. Plot
(b) is for the ``tightbinding'' scheme, where the potential $\Htbop$ is formed by making the barriers
at the boundaries of a module sufficiently thick to confine the wavefunctions within the module.
}
\label{fig:3levCB}
\end{figure}

\section{\label{sec:3lev}Current transport in a 3-level QCL}

An accurate estimation of the electrical transport characteristics of a QCL is quite
challenging, and more so for a terahertz QCL. Computationally intensive numerical
techniques based on Monte-Carlo simulations~\cite{callebaut:mcdm} and nonequilibrium Green's
functions~\cite{lee:neg2} have produced reasonably good results; however, an accurate
description of the temperature degradation mechanisms in terahertz QCLs is still lacking.  
Also, such computational techniques fall short of providing a good intuitive picture
of QCL operation. Hence, more often than not, analysis based on simple rate equations
are used during the design process. The 3-level model presented in this section
is a step toward developing a simpler transport model that still captures the nuances
of QCL operation and also provides a better understanding of how various parameters
affect its characteristics both below and above the lasing threshold.

For simplicity yet still captures the essentials, 
in what follows we discuss a 3-level QCL design with two wells per module
as shown in Fig.~\ref{fig:3levCB}~\cite{kumar:thesis1,kumar:twowell},
in which level $1'$ is the injector state, and the radiative transition is from level $3\rightarrow2$.
Even though typically QCL designs include many more levels per period, the results obtained
here hold for the triplet of levels consisting of the injector level and the two
radiative levels (levels $1',3~\rmand~2$ in Fig.~\ref{fig:3levCB}b), regardless of a particular design,
and including the mid-IR designs. Due to the various intrasubband and intersubband scattering mechanisms
that contribute to quantum transport, the electron wavepackets undergo dephasing, which manifests itself
as broadening of the energy levels. The tunnel-coupling between the levels of the same module is strong,
therefore, dephasing has a relatively negligible effect on the intramodule intersubband transport~\cite{callebaut:mcdm}.
However, the intermodule transport across the injector barrier is critically affected by the dephasing time
of the wavepackets since the $1'\rightarrow3$ tunneling time (half of the inter-well {\em Rabi}
oscillation period $\pi/\Omega_{1'3}$) is now longer and of the
same order. To incorporate the role of coherence in the
$1'\rightarrow3$ RT transport, we use density-matrix (DM) rate equations corresponding to
the tightbinding basis states of Fig.~\ref{fig:3levCB}(b)~\cite{callebaut:mcdm}. The time
evolution of the $3\times3$ DM for this basis set can be written as~\cite{kumar:thesis1}
\begin{eqnarray}
\ddt \rhomat_{(1',3,2)}
& = &
-\frac{i}{\hbar}
\left[ \Hextmat, \rhomat_{(1',3,2)} \right]
-\frac{i}{\hbar}
\left[ \bar{\bar{H}}', \rhomat_{(1',3,2)} \right]
\quad
\label{eq:dm3lev1}
\end{eqnarray}
where, $\Hmat'$ includes various radiative and non-radiative perturbation potentials
to the conduction-band potential $\Hextmat$. The advantage of writing the DMs
with the tightbinding basis of Fig.~\ref{fig:3levCB}(b) is two-fold. First, it allows
inclusion of dephasing phenomenologically by means of decay of the coherences associated
with levels $1'$ and $3$ due to non-zero off-diagonal terms in $\Hextmat$ corresponding
to those levels. Second, whereas the eigenstates of $\Hextop$ depend
sensitively on the externally applied electrical bias because of coupling between spatially
separated states, the chosen basis states
remain more or less independent of the bias in the range of interest (\ie\ close to
the ``design bias'' corresponding to the $1'-3$ alignment), which keeps the form-factors
for various intramodule scattering rate calculations approximately invariant of bias.
In a first order approximation, we assume RT to be independent of the electron wavevector
in the plane of the QWs. The subband lifetimes are assumed to be averaged over
the electron distribution within a subband and are to be calculated within
the Fermi's golden rule approximation. Expanding equation~\eqref{eq:dm3lev1},
we then obtain
\begin{eqnarray}
\ddt \rhomat_{(1',3,2)}
& = &
-i
\left[
\begin{pmatrix}
E_{1'}/\hbar & -\Omega_{1'3} & -\Omega_{1'2} \\
-\Omega_{1'3} & E_3/\hbar & 0 \\
-\Omega_{1'2} & 0 & E_2/\hbar \\
\end{pmatrix}
,
\rhomat_{(1',3,2)}
\right]
\nonumber \\ & &
\eightnegspc \eightnegspc \eightnegspc \eightnegspc 
\eightnegspc
+
\begin{pmatrix}
\frac{\rho_{33}}{\tau_{31}} + \frac{\rho_{22}}{\tau_{21}} & \frac{-\rho_{1'3}}{\tdgen{13}} & \frac{-\rho_{1'2}}{\tdgen{12}} \\
\frac{-\rho_{31'}}{\tdgen{13}} & -\frac{\rho_{33}}{\tau_3} - \frac{\rho_{33}-\rho_{22}}{\tst} & \frac{-\rho_{32}}{\tdgen{23}} \\
\frac{-\rho_{21'}}{\tdgen{12}} & \frac{-\rho_{23}}{\tdgen{23}} & \frac{\rho_{33}}{\tau_{32}} + \frac{\rho_{33}-\rho_{22}}{\tst} - \frac{\rho_{22}}{\tau_{21}}
\end{pmatrix}
\qquad
\label{eq:dm3lev2}
\end{eqnarray}
In the equation above, $\rhomat_{(1',3,2)} \equiv
\begin{pmatrix}
\rho_{1'1'} & \rho_{1'3} & \rho_{1'2}\\
\rho_{31'} & \rho_{33} & \rho_{32} \\
\rho_{21'} & \rho_{23} & \rho_{22} \\
\end{pmatrix}$
where $\rho_{ii}~(\equiv n_i)$ is taken as the number of electrons per module in level $i$ and
$\rho_{ij}$ is the coherence (also known as polarization) term for levels $i,j$.
In $\Hextmat$, the diagonal terms are the level energies given by
$\Hext_{nn}=\bra{n}\Hextop\ket{n}\approx E_n$ and the level anticrossings are represented
in the off-diagonal terms as $\Hext_{mn}=\bra{m}\Hextop-\Htbop\ket{n}\approx-\hbar\Omega_{mn}$.
For the lifetimes, $\tau_{ij}$ is the intersubband scattering time from $i\rightarrow j$,
$\frac{1}{\tdgen{ij}}\equiv\frac{1}{2\tau_i}+\frac{1}{2\tau_j}+\frac{1}{\Tpure}$ is the
dephasing rate for the coherence term $\rho_{ij}$ that consists of lifetime broadening
terms as well as a phenomenological broadening term $\Tpure$ due to interface-roughness
and impurity scattering~\cite{callebaut:mcdm}, and $\tst$ is due to the radiative stimulated
emission above the lasing threshold ($\rightarrow\infty$ below threshold).
The backscattering times $\tau_{23}$ and $\tau_{12}$ can also be included in
equation~\eqref{eq:dm3lev2} if relevant for a particular design. Note that $\Omega_{32}=0$ for
the chosen bases since subbands $3$ and $2$ are the eigenstates of the tightbinding potential
$\Htbop$ in Fig.~\ref{fig:3levCB}(b). By such a choice, the role of coherence
in $3\rightarrow2$ transport cannot be included. This does not introduce a large error in
estimating the current flow since $3$ and $2$ are strongly coupled. In other words, if
tightbinding bases were chosen for the radiative subbands $3$ and $2$ across the middle
``radiative'' barrier of the two-well module, $\Omega_{32}$ would be large and
the {\em Rabi} oscillation period $\pi/\Omega_{32}$ would be much smaller than the dephasing time
$\tdgen{32}$ for the $3\rightarrow2$ tunneling, making $\tdgen{32}$ inconsequential for the
$3\rightarrow2$ transport.

As a simpler alternative to more advanced methods, equation~\eqref{eq:dm3lev2} can model
some aspects of QCL transport fairly accurately. Although it could be solved
numerically, we seek analytical expressions that could provide a greater understanding about the
effect of various parameters on the transport. Toward that goal, we assume
$\Omega_{1'2}\approx0$ akin to a unity $1'\rightarrow3$ injection selectivity. This
assumption is in general valid for mid-IR designs due to a large radiative level
separation $E_{32}$ ($\equiv E_3-E_2$), and also becomes reasonable for the diagonal
terahertz designs~\cite{kumar:diagonal}. Within this approximation,
equation~\eqref{eq:dm3lev2} can be reduced to that with $2\times2$ matrices as follows
\begin{eqnarray}
\ddt
\rhomat_{(1',3)}
& = &
-i
\left[
\begin{pmatrix}
E_{1'}/\hbar & -\Omega_{1'3} \\
-\Omega_{1'3} & E_3/\hbar \\
\end{pmatrix}
,
\rhomat_{(1',3)}
\right]
\nonumber \\ & &
+
\begin{pmatrix}
\frac{\rho_{33}}{\tau_{31}}+\frac{\rho_{22}}{\tau_{21}} &
\frac{-\rho_{1'3}}{\td} \\
\frac{-\rho_{31'}}{\td} &
-\frac{\rho_{33}}{\tau_{31}}-\frac{\rho_{22}}{\tau_{21}} &
\end{pmatrix}
\qquad \qquad
\label{eq:dm3lev3}
\end{eqnarray}
where, $\rhomat_{(1',3)} \equiv
\begin{pmatrix}
\rho_{1'1'} & \rho_{1'3} \\
\rho_{31'} & \rho_{33} \\
\end{pmatrix}$, $\td\equiv\tdgen{13}$,
$\tau_3\equiv\frac{\tau_{31}\tau_{32}}{\tau_{31}+\tau_{32}}$, and
$\frac{\rho_{33}}{\tau_{3}}+\frac{\rho_{33}-\rho_{22}}{\tst}$
from equation~\eqref{eq:dm3lev2} is substituted by
$\frac{\rho_{33}}{\tau_{31}}+\frac{\rho_{22}}{\tau_{21}}$ in
equation~\eqref{eq:dm3lev3}, which holds when $\Omega_{1'2}=0$
as a statement of current continuity. Additionally, we can
write the following equation for $\rho_{22}$ below and above
the lasing threshold
\begin{eqnarray}
\rho_{22} =
\begin{cases}
\rho_{33} \displaystyle{\frac{\tau_{21}}{\tau_{32}}}
\qquad \qquad \;
\ldots \; \scriptstyle{(I<\Ith)}
\\
\rho_{33} - \delnth
\qquad \;
\ldots \; \scriptstyle{(I\geq\Ith)}
\end{cases}
\label{eq:dm3lev4}
\end{eqnarray}
where, $\delnth = (\rho_{33} - \rho_{22})_\rmth$ is the population inversion at threshold
that is assumed to remain constant beyond threshold. Equations~\eqref{eq:dm3lev3} and
\eqref{eq:dm3lev4} can be solved analytically for steady-state
$\left(\frac{d}{dt} \rightarrow 0\right)$.
With the constraint $\ntot=(\rho_{11} + \rho_{22} + \rho_{33})$, where $\ntot$ is the
total number of electrons per module and is a constant, the following expressions are
obtained for the current $I$~\cite{kumar:thesis1}
\begin{widetext}
\begin{eqnarray}
I & \equiv & 
|e|
\left( \frac{\rho_{33}}{\tau_{31}} + \frac{\rho_{22}}{\tau_{21}} \right)
=
\begin{cases}
\displaystyle{
|e|\ntot
\left[
\frac{2\Omsqgen{1'3} \td}
{4\Omsqgen{1'3}\td\left(\frac{\tau_{31}\tau_{32}}{\tau_{31}+\tau_{32}}\right)\left(1+\frac{\tau_{21}}{2\tau_{32}}\right)
+ \Delsqgen{1'3}\tdsq + 1}
\right]}
\quad \;\;\;
\ldots \; \scriptstyle{(I<\Ith)}
\\
\\
\displaystyle{
|e|\ntot
\left[
\frac{2\Omsqgen{1'3} \td \left(1 - \frac{2\delnth}{\ntot}\left[\frac{1-{\tau_{21}}/{(2\tau_{31})}}{1+{\tau_{21}}/{\tau_{31}}}\right]\right)}
{6\Omsqgen{1'3}\td\left(\frac{\tau_{31}\tau_{21}}{\tau_{31}+\tau_{21}}\right)+\Delsqgen{1'3}\tdsq + 1}
\right]}
\qquad \qquad
\qquad
\ldots \; \scriptstyle{(I\geq\Ith)}
\end{cases}
\eightnegspc
\label{eq:dm3levI}
\end{eqnarray}
\end{widetext}
where, $e$ is the unit charge and $\hbar\Delta_{1'3}\equiv E_{1'}-E_3$ is the energy detuning
between levels $1'$ and $3$ that is a function of the externally applied electrical bias. For
the case of ultra-short lifetime of the lower radiative state ($\tau_{21} \ll 2\tau_{32}$),
the below-threshold term in equation~\eqref{eq:dm3levI} becomes
\begin{eqnarray}
\fournegspc
I & = &
\displaystyle{
\frac{|e|\ntot}{2\tau_3}
\left(
\frac{4\Omsqgen{1'3} \td\tau_3}
{4\Omsqgen{1'3}\td\tau_3 + \Delsqgen{1'3}\tdsq + 1}
\right)}
\nonumber
\\ & & 
\qquad\qquad
\qquad\qquad
\quad
\ldots \; \scriptstyle{(\tau_{21}\ll 2\tau_{32}, \quad I<\Ith)}
\label{eq:dm2levI}
\end{eqnarray}
which reproduces the commonly used expression in literature~\cite{kazarinov:sl1,sirtori:tunnel}.

Since $\tau_{21} \ll 2\tau_{32}$ typically holds for a QCL design, the below-threshold term
for the 3-level model in equation~\eqref{eq:dm3levI} does not provide any new insight from
equation~\eqref{eq:dm2levI}. However,
the advantage of the model lies in describing transport above threshold. It
is instructive to write expressions for maximum current ($\Imax$), which flows at the $1'-3$
resonance ($\Delta_{1'3}=0$) for the case of a coherent RT process, \ie\ a large
injector coupling $\Omega_{1'3}$~\cite{sirtori:tunnel,eickemeyer:coherent}. The following
expressions are obtained for $\Imax$ and the level populations when threshold could not be attained
\begin{eqnarray}
\Imax & = &
\frac{|e|\ntot}{2\tau_3\left(1+\frac{\tau_{21}}{2\tau_{32}}\right)}
\nonumber
\\
n_3 & = & n_1 = \ntot\left(\frac{1}{2+\frac{\tau_{21}}{\tau_{32}}}\right)
\nonumber
\\
\Delta n & \equiv & n_3 - n_2 = \ntot\left(\frac{1-\frac{\tau_{21}}{\tau_{32}}}{2+\frac{\tau_{21}}{\tau_{32}}}\right)
\nonumber
\\ & &
\quad \
\ldots \;
\scriptstyle{\left(4\Omsqgen{1'3}\td\tau_3 \gg 1, \quad \Imax<\Ith\right)}
\label{eq:dm3levCoh1}
\end{eqnarray}

\begin{figure*}[htbp]
\centering
\includegraphics[width=5.50in]{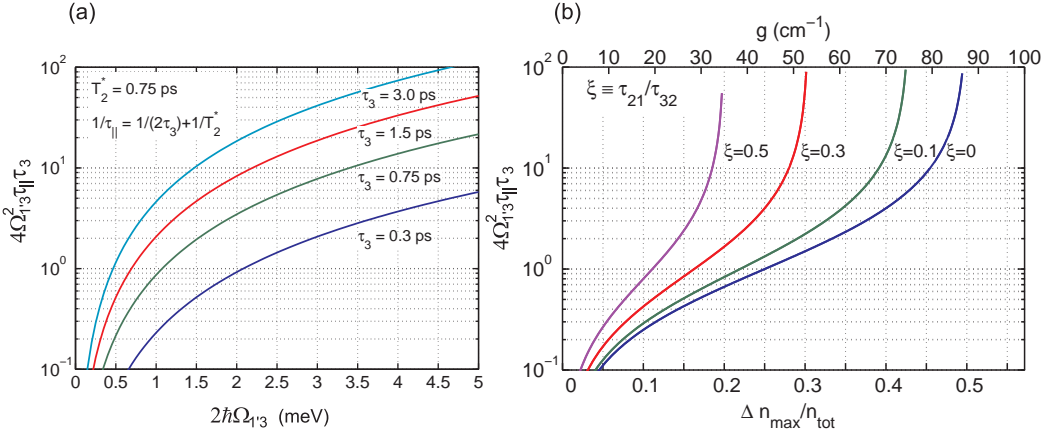}
\caption{
(a)~A plot of the factor $4\Omsqgen{1'3}\td\tau_3$ to determine the regime of RT transport through
the injector barrier ($\gg 1~$coherent, $\ll 1~$incoherent). A phenomenological value of
$\Tpure=0.75~\rmps$ is assumed. (b) A related plot to determine the maximum value of population
inversion ($n_3-n_2$) that can be attained for a given $\Omega_{1'3}$, calculated using
equation~\eqref{eq:3levDel13}. The top axis converts $\Delta n$ on the bottom axis to corresponding
values of peak gain for the following typical parameters:
$\ntot/\mathrm{Volume}=5\times10^{15}~\iiicm$, $f_{32}=0.5$, $\Delta\nu=1.0~\thz$, and using the
expression
$g \approx 70 \; \frac{[\Delta n / (10^{15} \ \mathrm{cm}^{-3})] \cdot f_{32}} {[\Delta \nu / (1 \ \mathrm{THz})]}\;
[\mathrm{cm}^{-1}]$ for GaAs material~\cite{kumar:review08} that assumes a Lorentzian gain
linewidth of $\Delta\nu$ (note that $f_{32}$ is the oscillator strength of the radiative
transition).
}
\label{fig:3levDel13}
\end{figure*}

\begin{figure*}[htbp]
\centering
\includegraphics[width=6.5in]{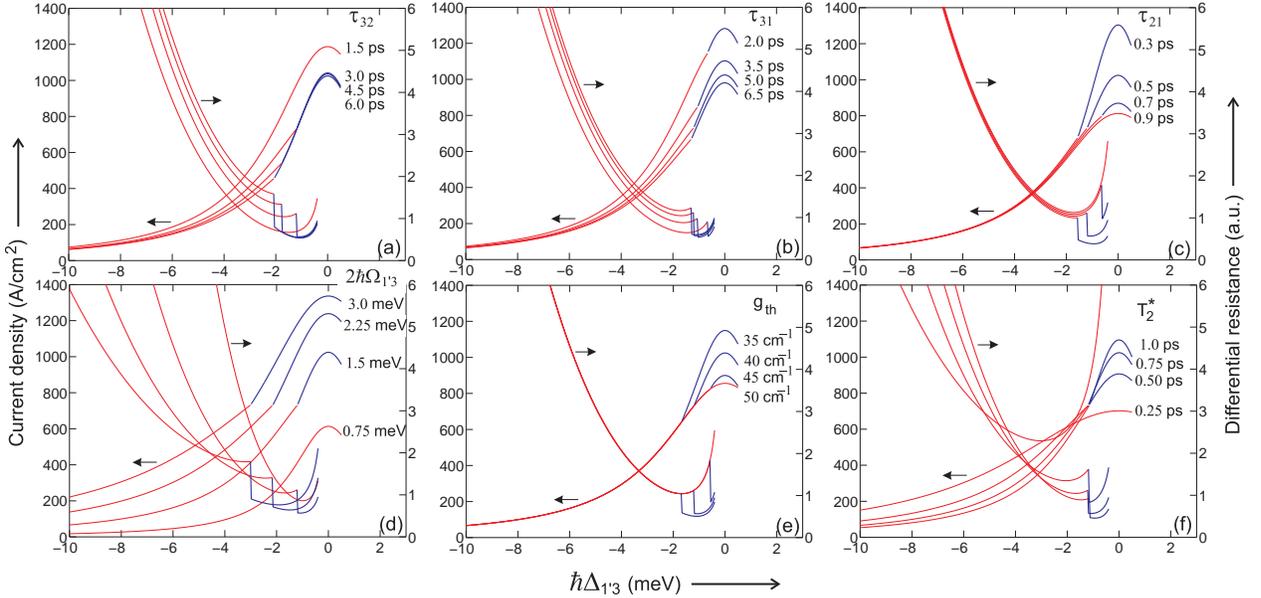}
\caption{
Calculations of the current density ($I/\mathrm{Area}$) as a function of the energy detuning $\hbar\Delta_{1'3}$
(which is proportional to the external bias voltage $V$, as in equation~\ref{eq:Vbias}), and its inverse slope
$\frac{\hbar}{|e|}\frac{d \Delta_{1'3}}{d I}$ (which is proportional to the differential resistance
$\Rdiff$, as in equation~\ref{eq:Rdiff}).
The calculations are done for a range of parameters using equation~\eqref{eq:dm3levI} for the 3-level
model. Following default values are chosen corresponding to the typical values for RPTQCL
designs: $2\hbar\Omega_{1'3}=1.5~\meV$, $\Tpure=0.75~\rmps$, $\gth=40/\mathrm{cm}$~\cite{RTpaperf1:footnote},
$\tau_{21}=0.5~\rmps$, $\tau_{32}=3.0~\rmps$, $\tau_{31}=5.0~\rmps$. A doping density
$\ntot/\mathrm{Volume} = 5\times10^{15}~\iiicm$ is used. The radiative transition is assumed to
have a Lorentzian linewidth of $\Delta \nu = 1.5~\thz$ (the broad linewidth is based on the findings
of section~\ref{sec:4lev}) and an oscillator strength of $f_{32}=0.6$ to determine the population
inversion $\delnth$ required to attain a particular value of threshold gain $\gth$. The thin (red)
portion of the curves corresponds to $I<\Ith$, while the thick (blue) region is for $I>\Ith$ where
the occurrence of threshold is marked by a discontinuity in $\Rdiff$. Each of
the subplots show the variation of the \IVs\ and the \RVs\ when only a single parameter is changed,
the others being kept at the values mentioned above. Note that additional linewidth broadening due
to coherent effects (discussed in section~\ref{sec:4lev}) is not considered for calculations
of the curves above threshold.
}
\label{fig:3levIVs}
\end{figure*}

The results in equation~\eqref{eq:dm3levCoh1} are slightly different than that obtained
using a two-level model in Ref.~\onlinecite{sirtori:tunnel} as a consequence of adding an additional
level (\ie\ the lower laser subband $2$) in these calculations. However, the commonly used two-level
model result $\Imax=|e|\ntot/(2\tau_3)$ is recovered in the limit $\tau_{21}\ll\tau_{32}$
as it is to be expected. Equation~\eqref{eq:dm3levCoh1} suggests that $\Imax$ for coherent injection
is independent of $\Omega_{1'3}$, and therefore the thickness of the injector barrier, and is
limited by the upper state lifetime $\tau_3$. Typical values of $\Omega_{1'3}$ that bring a QCL
within this limit could be determined from Fig.~\ref{fig:3levDel13}(a).
The value of $\Imax$, however, is different if enough population inversion could be attained
prior to the $1'-3$ resonance and the device starts lasing, in which case stimulated
emission lowers the upper state lifetime. In such a scenario, the maximum current is limited
by the lower state lifetime $\tau_{21}$ instead, and is given by
\begin{eqnarray}
\Imax & = &
\frac{|e|\ntot}{3\tau_{21}}
\left[ 1-\frac{2\delnth}{\ntot} +\frac{\tau_{21}}{\tau_{31}}\left(1+\frac{\delnth}{\ntot}\right) \right]
\nonumber
\\
n_3 & = & n_1 = \left(\frac{\ntot}{3} + \frac{\delnth}{3}\right)
\nonumber
\\ & &
\quad \
\ldots \;
\scriptstyle{\left(6\Omsqgen{1'3}\td\tau_{21} \gg 1, \quad \Imax>\Ith\right)}
\label{eq:dm3levCoh2}
\end{eqnarray}
Equation \eqref{eq:dm3levCoh2} implies that a QCL with a large injector coupling
$\Omega_{1'3}$ and a short lower state lifetime $\tau_{21}$ can possibly obtain a 
large dynamic range in current due to a large $\Imax$ (see Figs.~\ref{fig:3levIVs}c 
and \ref{fig:3levIVs}d). However, it may be noted that
for coherent injection, the gain linewidth of the QCL becomes additionally
broadened as the injector coupling $\Omega_{1'3}$ is increased.
Consequently, a larger $\delnth$ is required to attain a certain value
of peak gain, which mitigates the increase in the value of $\Imax$. This will become
more clear in section~\ref{sec:4lev} from the gain spectrum calculations for some
specific cases.

Approaching threshold, the value of current that must flow through the QCL structure to establish
a population inversion of $\delnth$ is calculated to be
\begin{equation}
\Ith=\frac{|e|\delnth}{\tau_3\left(1-\frac{\tau_{21}}{\tau_{32}}\right)}
\label{eq:3levIth}
\end{equation}
While the above expression for $\Ith$ can also be derived from a simple rate equation
analysis, the minimum value of $\Omega_{1'3}$ needed for a current $\Ith$ 
to flow through the structure is derived from the present 3-level DM model as
\begin{equation}
4\Omsqgen{1'3}\td\tau_3 >
\frac{2\delnth}{\ntot - 2 \delnth - (\ntot + \delnth)\frac{\tau_{21}}{\tau_{32}}}
\label{eq:3levDel13}
\end{equation}
This result is obtained from equation~\eqref{eq:dm3levI} as a necessary
condition for the two expressions to have the same value for some particular
bias $\Delta_{1'3}$. An estimate of the minimum value of $\Omega_{1'3}$ required to
meet this condition can be obtained from the plots in Fig.~\ref{fig:3levDel13}(b).
Figure~\ref{fig:3levDel13}(b) can also be interpreted in a different way, since it
determines the maximum value of population inversion that can be attained for a given
$\Omega_{1'3}$. For example, in the limit $\tau_{21}\rightarrow0$, a maximum of
$\Delta n=\ntot/2$ can be obtained if $4\Omsqgen{1'3}\td\tau_3\gg 1$.

\subsection{\label{subsec:3levIVs}Current-voltage characteristics for typical parameters}

To gauge the effect of various parameters on QCL's performance, Fig.~\ref{fig:3levIVs}
shows calculation of the current-voltage (\IV) and differential-resistance-voltage
(\RV, $\Rdiff=\frac{d V}{d I}$)
characteristics for a range of parameters using the analytical expressions in
equation~\eqref{eq:dm3levI}. The \IVs\ show a discontinuity in slope at the occurrence
of the lasing threshold, as was first discussed in detail in Ref.~\onlinecite{sirtori:tunnel}.
However, any discussion about the differential resistance in the present context is left
until later. For the \IVs\ the main features to be noted are the values of the threshold
current density $\Jth$ and the maximum current density $\Jmax$. In general
for any design, the goal is to obtain a larger dynamic range for lasing $\Jmax/\Jth$
while also keeping the value of $\Jth$ low. A larger dynamic range leads to greater
amount of the optical power output, and also a higher operating temperature $\Tmax$.
As it is to be expected, the plots show that the laser performance improves by making
$\tau_{31}$ and $\tau_{32}$ larger, and $\tau_{21}$ smaller.

\begin{figure}[htbp]
\centering
\includegraphics[width=3.4in]{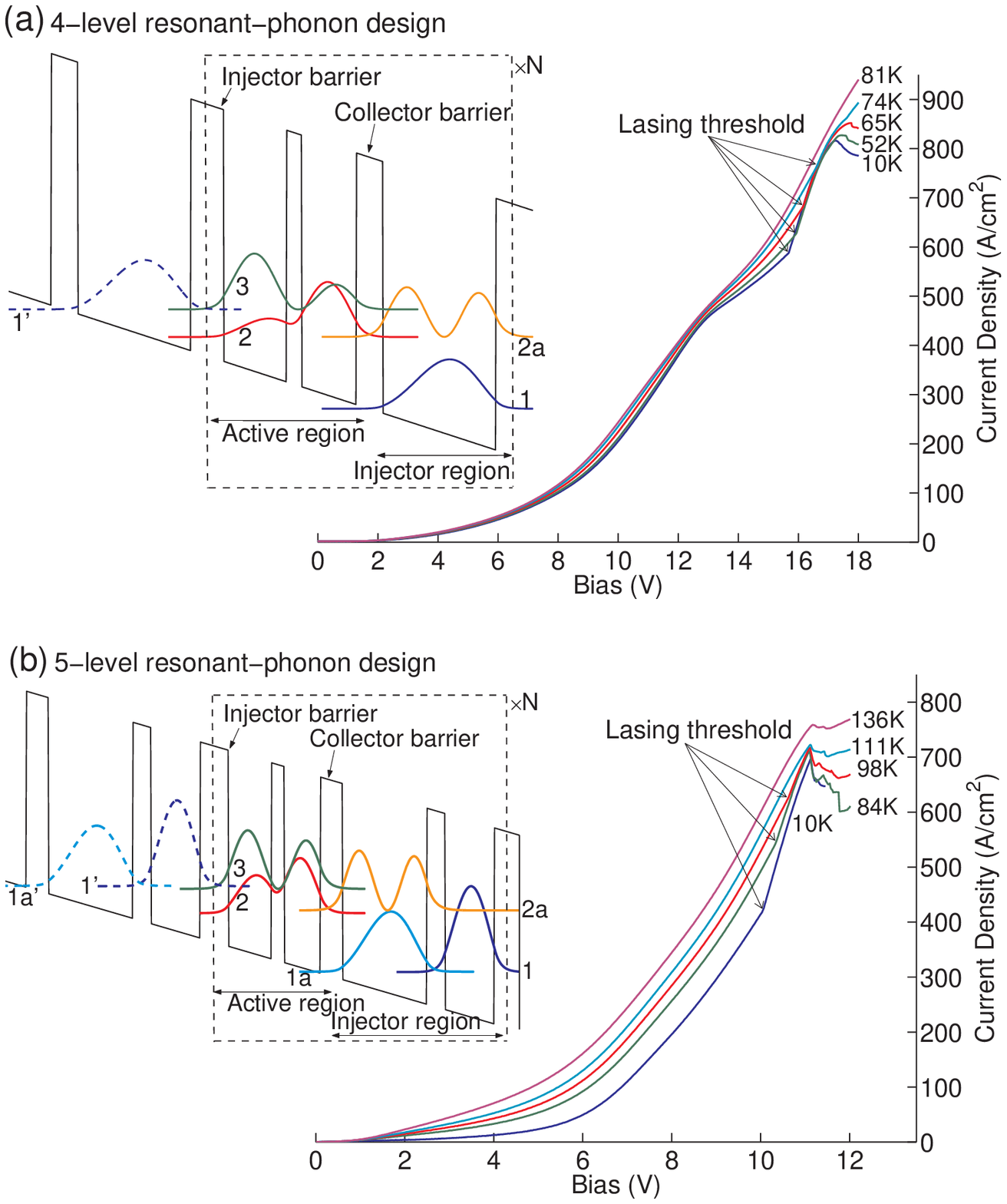}
\caption{
Typical experimental \IVs\ versus temperature for (a)~a 4-level, and
(b)~a 5-level resonant-phonon terahertz QCL, respectively.
The corresponding band diagrams show tightbinding wavefunctions at design-bias calculated by
splitting the QCL module into multiple submodules across the relevant barriers. The radiative
transition is from $3\rightarrow2$, and the depopulation of the lower
level is via $2\rightarrow2a$ RT and $2a\rightarrow1/1a$ electron-longitudinal-optical
phonon scattering where $E_{21}\approx\hwLO$.
The \IVs\ are measured in continuous-wave (cw) operation for metal-metal
ridge lasers. The plot in (a) is from a $3.9~\thz$ QCL labeled OWI222G~\cite{kumar:diagonal} of dimensions
$60~\um\times310~\um$ ($\Tmaxc\sim76~\rmK$), and that in (b) is from a $2.7~\thz$ QCL labeled
FL178C-M10~\cite{kumar:review08} of dimensions $35~\um\times670~\um$ ($\Tmaxc\sim108~\rmK$).
The onset of lasing results in a change in the slope of the \IV, which is indicated by arrows for
the curves recorded below the $\Tmaxc$.
}
\label{fig:RPCBIVs}
\end{figure}

We now compare the 3-level model \IVs\ in Fig.~\ref{fig:3levIVs} to those experimentally
measured for two different RPTQCL designs. Fig.~\ref{fig:RPCBIVs} shows typical results
from a 4-level~\cite{kumar:owi,luo:threewell,kumar:diagonal} and a 5-level~\cite{williams:laser,kumar:review08}
RPTQCL respectively. The structures in Fig.~\ref{fig:RPCBIVs}(a) and \ref{fig:RPCBIVs}(b)
can be qualitatively analyzed with the 3-level model of this section by using the following
expression for the effective lifetime of the lower laser level $2$, which is
depopulated through RT into $2a$, and subsequently via electron-longitudinal-optical (e-LO)
phonon scattering into the injector level(s) (\ie\ the {\em resonant-phonon} scheme).
The following expression can be derived using a 2-level density-matrix for levels
$2$ and $2a$ exclusively and assuming $2\rightarrow2a$ tunneling to be the only mechanism
for carrier injection into level $2a$~\cite{kumar:thesis1} 
\begin{equation}
\tau_{2,\mathrm{eff}}
=
\tau_{2a}
\left(
\frac
{2\Omsqgen{2,2a}\tdgen{}\tau_{2a} + \Delsqgen{2,2a}\tdsqgen{} + 1}
{2\Omsqgen{2,2a} \tdgen{}\tau_{2a}}
\right)
\label{eq:t2eff}
\end{equation}
\begin{figure}[htbp]
\centering
\includegraphics[width=2.5in]{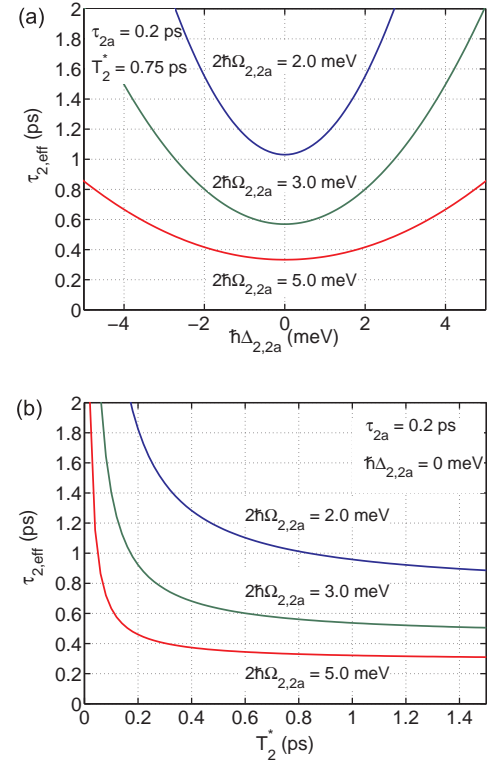}
\caption{Plots of the effective lower-level lifetime $\tau_{2,\mathrm{eff}}$ (also equivalent to the
{\em tunneling time} through the collector barrier) for the resonant-phonon
designs of the type shown in Fig.~\ref{fig:RPCBIVs} calculated using equation~\eqref{eq:t2eff}. Plot
(a) shows $\tau_{2,\mathrm{eff}}$ versus the energy detuning $\hbar\Delta_{2,2a}$ between the levels $2$
and $2a$ and (b) shows $\tau_{2,\mathrm{eff}}$ versus the phenomenological dephasing time $\Tpure$ calculated
at the $2-2a$ resonance bias ($\hbar\Delta_{2,2a}=0$). $\tau_{2a}\sim0.2~\ps$ is assumed owing to the
typical value obtained in GaAs/\AGAf\ quantum wells for $E_{2a,1}\sim\hwLO$.
}
\label{fig:t2eff}
\end{figure}

The effective lower-level lifetime $\tau_{2,\mathrm{eff}}$, which is also equivalent to the {\em tunneling
time} through the collector barrier, is plotted for three different values of the
collector anticrossing $2\hbar\Omega_{2,2a}$ in Fig.~\ref{fig:t2eff}. The best performing RPTQCLs have large
collector anticrossings in the range of $2\hbar\Omega_{2,2a}\sim4-5~\meV$, in which case $\tau_{2,\mathrm{eff}}$
varies little in the operating bias range close to the $2-2a$ resonance ($\hbar\Delta_{2,2a}=0$). For example,
$2\hbar\Omega_{2,2a}\sim4.7~\meV$ for the 4-level QCL in Ref.~\cite{kumar:diagonal} and the energy detuning
$\hbar\Delta_{2,2a}$ approximately varies from $-2~\meV$ to $2.5~\meV$ in operating bias range of the design.
As seen from Fig.~\ref{fig:t2eff}(a), $\tau_{2,\mathrm{eff}}$ changes little within such a range of detuning.

We can similarly take the combined population $\ntot=n_{1'}+n_2+n_3$ to be approximately bias independent.
The population of level $2a$ is likely to be negligible due to its very short lifetime $\tau_{2a}\sim0.2~\ps$,
and that of level $1a$ in Fig.~\ref{fig:RPCBIVs}(b) is likely to vary negligibly with bias in the operating bias
range of interest due to the typically
large values of the intra-injector anticrossing $2\hbar\Omega_{1a,1}\sim4-5~\meV$ in the 5-level design
Fig.~\ref{fig:RPCBIVs}(b).
Given the bias independence of $\ntot=n_{1'}+n_2+n_3$ and the lower-level lifetime $\tau_{2,\mathrm{eff}}$ for
the resonant-phonon designs in Fig.~\ref{fig:RPCBIVs}, we can now qualitatively compare the 3-level model \IVs\
in Fig.~\ref{fig:3levIVs} to the experimental continuous-wave (cw) \IVs\ in Fig.~\ref{fig:RPCBIVs}. For the latter,
it can be noticed that $\Jmax$, which is characterized by the occurrence of a negative-differential resistance (NDR)
region in the \IV, changes little up until the maximum operating temperature of the laser $\Tmaxc$.
However, beyond $\Tmaxc$, the value of $\Jmax$ increases with temperature.
This characteristic is best described by the calculated \IVs\ in Fig.~\ref{fig:3levIVs}(a)
since the above-threshold current expression in equation~\eqref{eq:dm3levI} is independent
of the upper state to lower state lifetime $\tau_{32}$ to a first order. Hence, it is
likely that one of the dominant temperature degradation mechanism for RPTQCLs
is a reduction in the upper state ($u$) to lower state ($l$) lifetime, which is
mostly attributed to the thermally activated LO phonon scattering from the upper
state~\cite{williams:review,kumar:review08}.
The recently developed terahertz QCL design with a very diagonal radiative transition
is based on this observation to effectively increase the $u\rightarrow l$ lifetime at
high temperatures~\cite{kumar:diagonal}. Additionally, it seems less likely that
an increase in the effective lifetime of the lower state (potentially due to $1\rightarrow2$
thermal backfilling or increased dephasing for the $2\rightarrow2a$ RT that affects
$\tau_{2,\mathrm{eff}}$ as in Fig.~\ref{fig:t2eff}b),
or a broadening of the gain linewidth~\cite{nelander:broadening} are more dominant
temperature degradation mechanisms than that mentioned above; otherwise, the observed
experimental behavior would have corresponded more closely with Figs.~\ref{fig:3levIVs}(e)
or \ref{fig:3levIVs}(c), respectively. That is, the peak current density $\Jmax$ would
decrease at elevated temperatures.

\begin{figure}[htbp]
\centering
\includegraphics[width=3.0in]{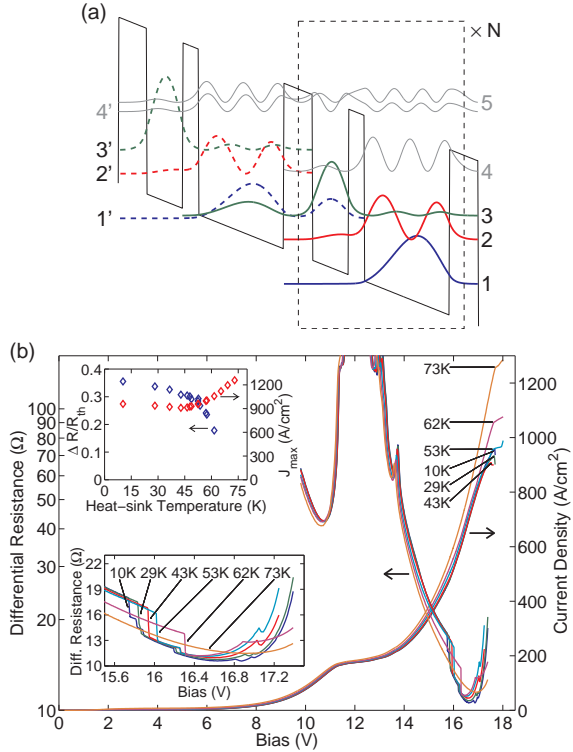}
\caption{
(a)~Conduction band diagram of a recently demonstrated two-well terahertz QCL design, labeled
TW246~\cite{kumar:twowell}. Only three levels ($1,2$ and $3$) participate in
transport at low temperatures; however, at higher temperatures additional parasitic levels
($4$ and $5$) are also likely to contribute to the current flow.
(b)~Experimental cw \IVs\ and \RVs\ at different heat-sink temperatures from a $20~\um\times1.56~\rmmm$
metal-metal ridge laser with a $\Tmaxc\sim63~\rmK$. The lasing threshold is marked by a
discontinuity in the $\Rdiff$ curves that is shown more clearly in an expanded view in the lower
inset. The upper inset shows the variation of fractional discontinuity in $\Rdiff$ at threshold
$\drrfrac\left(\equiv\frac{\Rdiffm-\Rdiffp}{\Rdiffm}\right)$ and of $\Jmax$ with temperature.
}
\label{fig:twowell}
\end{figure}

The two-well design of Fig.~\ref{fig:3levCB}, which features an {\em intrawell-phonon}
(IP) depopulation scheme as opposed to the RP depopulation scheme of Fig.~\ref{fig:RPCBIVs},
 has also been experimentally realized recently~\cite{kumar:twowell}. The design-bias band
diagram of the realized structure is shown in Fig.~\ref{fig:twowell}(a) including higher
energy parasitic levels that are believed to contribute to electron transport at elevated
temperatures. The experimental \IVs\ and \RVs\ from a representative ridge laser in cw
operation are shown in Fig.~\ref{fig:twowell}(b). The temperature variation of the
\IVs\ shows that $\Jmax$ for this IP QCL decreases slightly with temperature up to
$T\sim45~\rmK$ and subsequently increases steeply with temperature until and beyond the
$\Tmaxc\sim63~\rmK$. This unique dependence of $\Jmax$ with $T$, also shown explicitly
in the inset of Fig.~\ref{fig:twowell}(b), is different from that of the RP QCLs
in Fig.~\ref{fig:RPCBIVs}, and is attributed to the onset of a temperature degradation
mechanism that is in addition to the reduction in $\tau_{32}$ due to thermally activated
e-LO phonon scattering from $3\rightarrow2$. It is postulated that the lifetime of the
upper state $3$ decreases additionally due to absorption of non-equilibrium (hot) LO phonons
via $3'\rightarrow(4',5)$ scattering~\cite{kumar:twowell}. Any such scattered electrons
quickly relax back eventually into the injector level of a neighboring module due to the
short lifetime of the parasitic levels $4$ and $5$ ($\tau_{4'}\sim\tau_{5}\sim0.3~\ps$). Such
a leakage mechanism can be considered approximately as an effective reduction in $\tau_{31}$
in order to apply the results of the 3-level model to explain the behavior in
Fig.~\ref{fig:twowell}(b), which then becomes consistent with calculated \IV\ behavior in
Figs.~\ref{fig:3levIVs}(b). Note that the upper laser state in the RP designs
is spatially isolated from the wide injector well(s) and hence from the higher energy
parasitic levels~\cite{kumar:review08}. Hence, the aforementioned hot-phonon
mediated leakage mechanism is weaker for the RPTQCLs as compared to the IP QCL
structure of Fig.~\ref{fig:twowell}(a).

\subsection{\label{subsec:3levR}Discontinuity in differential resistance at threshold}

\begin{figure}[htbp]
\centering
\includegraphics[width=2.5in]{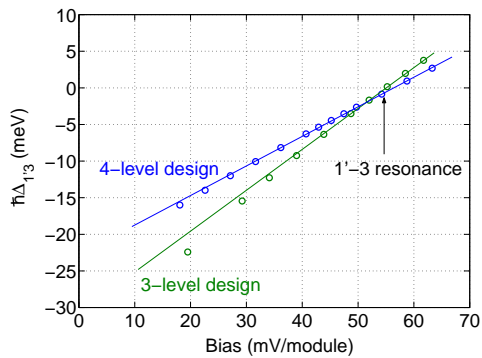}
\caption{Detuning energy $\hbar\Delta_{1'3}$ as a function of the voltage 
per QCL module ($=V/\Np$ in equation~\ref{eq:Vbias}) for the tightbinding
wavefunctions of the 4-level resonant-phonon design of Fig.~\ref{fig:RPCBIVs}(a) and the
3-level intrawell-phonon design of Fig.~\ref{fig:twowell}(a) respectively. Small circles
represent calculated values, which are overlaid with straight line fits to the data points
close to the $1'-3$ resonance. The fits indicate that $\hbar\Delta_{1'3}$ is linearly related
to $V/\Np$ close to the $1'-3$ resonance (\ie\ $f_V$ in equation~\ref{eq:Vbias}
is a constant). Typically, lasing threshold occurs after the $1'-2$ resonance ($\sim30~$mV/module)
beyond which the current continues to increase up to the $1'-3$ resonance when $\Jmax$ occurs.
}
\label{fig:fV}
\end{figure}

The onset of lasing in a QCL is characterized by a slope-discontinuity in its \IV\
characteristics~\cite{sirtori:tunnel}. This is a manifestation of the increased rate of carrier
flow due to the stimulated emission process from the upper to the lower state,
such that the population inversion is kept constant beyond threshold. The relative change
in the differential resistance $\Rdiff=dV/dI$ of the device at threshold is a useful parameter
since it is easily and accurately measurable, and is shown below to be directly proportional
to the value of $\delnth$. We can relate the externally applied bias voltage across all the 
periods of the QCL to the energy detuning $\hbar\Delta_{1'3}$ as 
\begin{equation}
V =
\frac{\hbar\Delta_{1'3}\Np}{f_V|e|}
\label{eq:Vbias}
\end{equation}
where, $\Np$ is the number of repeated periods in the QCL structure, and $f_V$ is the
fraction of per module voltage bias $V/\Np$ that appears as the energy difference
between levels $1'$ and $3$. $f_V$ is typically a slowly varying non-linear function
of the applied voltage $V$ and is a characteristic of a particular design as shown in
Fig.~\ref{fig:fV}. For the main result presented in this section, the particular form
of $f_V$ is inconsequential and it is assumed to be a constant in the bias range right below
and above the threshold. Using equation~\eqref{eq:Vbias} the differential resistance $\Rdiff$
becomes
\begin{equation}
\Rdiff
= \frac{d V}{d I}
\approx \frac{\hbar\Np}{f_V|e|} \frac{d E_{1'3}}{d I}
\label{eq:Rdiff}
\end{equation}

In equation~\eqref{eq:Rdiff}, we have dropped the term corresponding to $\frac{d f_V}{d I}$
since $f_V$ is approximately a constant as can be seen from Fig.~\ref{fig:fV}.
The calculations in Fig.~\ref{fig:fV} are done without accounting for band-bending due to
dopant-carrier segregation, which is typically insignificant due to the low-doping in such QCL
structures~\cite{williams:thesis}.
The value of $\frac{d f_V}{d I}$ is likely to be negligible in comparison to the leading term
of equation~\eqref{eq:Rdiff} even if band-bending were found to be significant for a particular
design. We can now derive analytical expressions for the ratio as well as the fractional change
in differential resistance just below ($\Rdiffm$) and above ($\Rdiffp$) threshold using
equation~\eqref{eq:dm3levI} to calculate $\frac{d\Delta_{1'3}}{dI}$. The individual expressions
for $\Rdiffm$ and $\Rdiffp$ are complicated; however, the ratio $\frac{\Rdiffp}{\Rdiffm}$
and hence the fractional change in $\Rdiff$ at threshold $\drrfrac$ is expressed
in a surprisingly concise form as~\cite{kumar:thesis1}
\begin{eqnarray}
\frac{\Rdiffp}{\Rdiffm} & = &
\left[
1-
\frac{2\delnth}{\ntot}
\left(\frac{1-\frac{\tau_{21}}{2\tau_{31}}}{1+\frac{\tau_{21}}{\tau_{31}}}\right)
\right]
\label{eq:drr1}
\\
\drrfrac & \equiv &
\frac{\Rdiffm-\Rdiffp}{\Rdiffm}
=
\frac{2\delnth}{\ntot}
\left(\frac{1-\frac{\tau_{21}}{2\tau_{31}}}{1+\frac{\tau_{21}}{\tau_{31}}}\right)
\label{eq:drr2}
\end{eqnarray}

Note that equation~\eqref{eq:drr2} is derived for the case of unity injection efficiency,
\ie\ $\Omega_{1'2}\approx 0$ in the $3\times3$ DM equation~\eqref{eq:dm3lev2}. 
Hence, within the approximation that the current
flow due to the $1'\rightarrow2$ channel is negligible in comparison to the $1'\rightarrow3$
channel {\em at threshold}, the expression derived for $\drrfrac$ holds true irrespective of
the nature of the RT transport (coherent or incoherent) across the injector barrier.
At low temperatures, $\tau_{21}\ll\tau_{31}$, and hence $\drrfrac\approx\frac{2\delnth}{\ntot}$
gives an absolute measurement of the population inversion in the laser as a fraction of the
combined populations of the injector level $1'$ and the laser levels $3$ and $2$
($\ntot\equiv n_{1'}+n_2+n_3$).

The slope efficiency of a QCL's optical power output is written as
\begin{equation}
\frac{d \Pout}{d I} =
\frac{\Np \hwz }{|e|}
\frac{\alpha_\rmm}{(\alpha_\rmw + \alpha_\rmm)}
\eta
\label{eq:slopeeffthresh2}
\end{equation}
where $\omega_0$ is the lasing frequency, $\alpha_\rmm$ is the radiative (mirror) loss in
the cavity, $\alpha_\rmw$ is the material (waveguide) loss, and  $\eta$ is the internal
quantum efficiency of the QCL structure. $\eta<1$ due to the fact that
$n_3$, and hence the non-radiative component of the current $I$ ($=|e| n_3/\tau_3$),
continues to increase with $I$ above threshold (even as $\Delta n=n_3-n_2$
remains fixed at $\delnth$), which causes the current above threshold
$(I-\Ith)$ to be not entirely due to radiative transitions. An expression for
$\eta$ is derived as
\begin{eqnarray}
\eightnegspc
\eta
=\frac{1-\frac{\tau_{21}}{\tau_{32}}}{1+\frac{\tau_{21}}{\tau_{31}}}
=1-\frac{\frac{d n_3}{dI}|_{I=\Ithp}}{\frac{d n_3}{dI}|_{I=\Ithm}}
\label{eq:iqe1}
\end{eqnarray}
where,
\begin{eqnarray}
\eightnegspc
\frac{d n_3}{dI}|_{I=\Ithp}
& = &
\frac{1}{|e|}\frac{\tau_{21}\tau_{31}}{(\tau_{21}+\tau_{31})}
\nonumber
\\
\frac{d n_3}{dI}|_{I=\Ithm}
& = &
\frac{1}{|e|}\tau_3
\label{eq:iqe2}
\end{eqnarray}
Equation~\eqref{eq:iqe1} for $\eta$ can also be derived from a rate equation model
since $n_3$ can be expressed in terms of $I$ regardless of the nature of
$1'\rightarrow3$ RT transport. In literature~\cite{sirtori:tunnel,ajili:highpower},
$\drrfrac$ has similarly been derived from a rate equations approach, arguing that
$n_3\propto \Delta_{1'3}$, and hence the applied voltage $V$, in which case $\drrfrac$
becomes same as $\eta$ (for unity injection efficiency), and hence an indicator
of the ratio of laser level lifetimes $\tau_{21}/\tau_{32}$.
However, the expression for $\drrfrac$ in equation~\eqref{eq:drr2} is starkly different
from that of $\eta$ in equation~\eqref{eq:iqe1}. While $\eta\rightarrow 1$ for
$\tau_{21}\rightarrow 0$, but $\drrfrac\rightarrow \frac{2\delnth}{\ntot}$.
Also, $\eta$ strongly depends on the upper state to lower state lifetime $\tau_{32}$,
whereas $\drrfrac$ is independent of $\tau_{32}$. Hence, the result derived with the
3-level DM model has important implications in the way the slope discontinuity is
interpreted for the \IV\ of a QCL at threshold. In essence, $\drrfrac$ is an indicator
of the population inversion at threshold and not the ratio of level lifetimes, which
are two different aspects of laser operation.

\begin{figure}[htbp]
\centering
\includegraphics[width=3.3in]{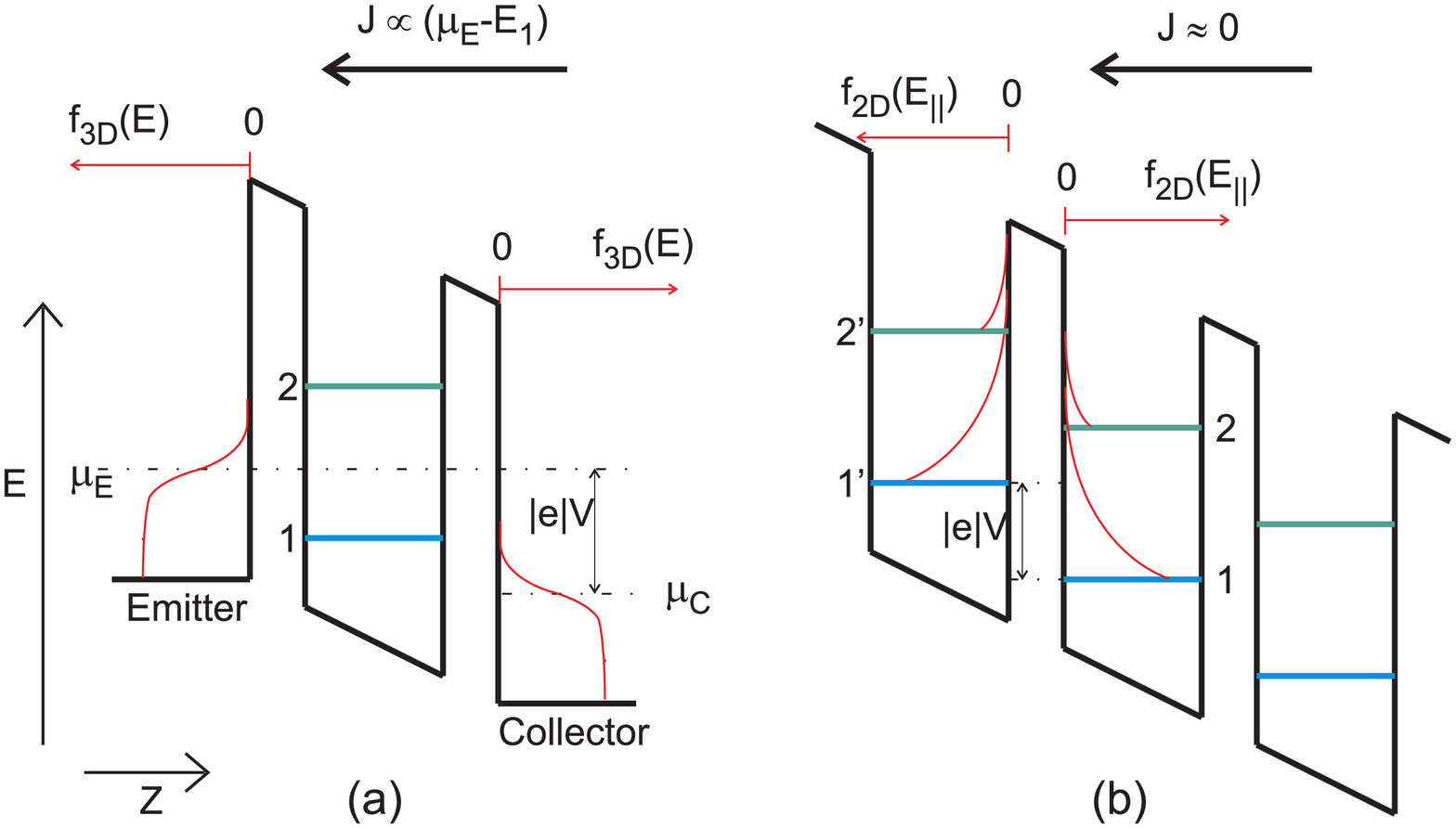}
\caption{
Illustrative band diagrams for (a) double-barrier resonant-tunneling diode structure~\cite{chang:rt}
that consists of a single quantum-well sandwiched between degenerately doped emitter and collector
regions, and (b) a superlattice structure~\cite{esaki:sl} that consists of multiple
repeated quantum-wells of which only three wells are shown. In (a), the shown Fermi distributions
(indicated by thin red lines) for the carrier populations apply to a quasi-continuous carrier distribution
in three-dimensions, whereas in (b), the shown carrier distribution applies only to the two-dimensional
momentum space in the $x-y$ plane (since the carriers are confined in the $z$ direction), and is typically
Maxwell-Boltzmann like (\ie\ the Fermi energy $\mu_n$ lies much below the energy of the bottom of the
subband $E_n$ indicated by thick horizontal lines). In (a), $V$ is the voltage applied between the
emitter and the collector that appears as the difference between the respective Fermi energies, and in
(b) $V$ represents the voltage per repeated module of the superlattice that appears as the difference
in the energy of the bottom of the subbands.
}
\label{fig:rtunneling}
\end{figure}

By means of Fig.~\ref{fig:rtunneling}, we now show that the assumption $n_3\propto\Delta_{1'3}$
which is the basis of the previously held belief of $\drrfrac$ being equivalent to
$\eta$, is incorrect. For the conventional resonant-tunneling diode (RTD) structures,
carriers tunnel from the three-dimensional states in the emitter into the quasi-two-dimensional
states in the well. Considering only first-order tunneling processes (\ie\ conservation
of in-plane momentum in tunneling through the barriers), the current flow is proportional
to the Fermi energy $\mu_\mathrm{E}$ in the emitter and hence the applied voltage $V$ (as
shown in Fig.~\ref{fig:rtunneling}(a), see Ref.~\onlinecite{ohno:rtd} for example). In such
a case, the population $n_1$ of the subband~$1$ in Fig.~\ref{fig:rtunneling}(a) will
indeed be proportional to $V$. However, for a semiconductor superlattice structure of
which QCLs are specific examples, the current flow will be negligible if the
subbands are significantly off-resonance in the neighboring quantum-wells even
if the carrier distributions overlap in energy. This is because total energy cannot 
be conserved in such a tunneling process since the in-plane momentum conservation
needs to be satisfied. The applied bias field is in the $z$ direction and hence
affects the bottom energy of the subbands and their $z$-wavefunctions, however, the
quasi-Fermi levels within the respective subbands represent the in-plane carrier
distribution and are not directly affected by the applied voltage $V$. As a result,
the subband populations cannot be assumed to be directly proportional to $V$. Instead,
the subband populations and their thermal distributions (and hence the respective
quasi-Fermi levels) are determined by the
various intersubband and intrasubband scattering processes and also by the
resonant-tunneling current flow induced by the coupling of the subband
wavefunctions across the barriers as determined by the anticrossing energies
$2\hbar\Omega_{ij}$ and detuning energies $\hbar\Delta_{ij}$. Note that higher
order corrections to the resonant-tunneling current flow that relax the
requirement of in-plane momentum conservation are not included in
this simplified picture~\cite{wacker:transport,willenberg:bloch}.

\begin{figure}[htbp]
\centering
\includegraphics[width=2.5in]{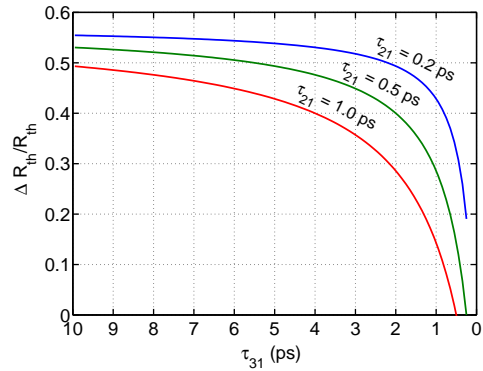}
\caption{
Variation of the fractional change in differential resistance at threshold $\drrfrac$
with $\tau_{31}$ calculated using the expression in equation~\eqref{eq:drr2} with
same typical values of different parameters as in Fig.~\ref{fig:3levIVs}.  
}
\label{fig:drr}
\end{figure}

Figure~\ref{fig:drr} shows typical variation of $\drrfrac$ with
$\tau_{31}$ according to equation~\eqref{eq:drr2}, which shows that $\drrfrac$
decreases rapidly as $\tau_{31}$ becomes smaller and approaches $\tau_{21}$.
The trend in Fig.~\ref{fig:drr} could be compared to the experimental variation
of $\drrfrac$ with temperature for the two-well QCL shown in the upper inset
of Fig.~\ref{fig:twowell}(b), which indicates a steady degradation of
its effective $\tau_{31}$ with temperature due to a predicted hot-phonon
effect~\cite{kumar:twowell}. Also note that $\Jmax$ and $\drrfrac$ in
Fig.~\ref{fig:twowell}(b) have a correlated behavior where $\Jmax\uparrow$
as $\drrfrac\downarrow$. This validates the expression in
equation~\eqref{eq:dm3levI} for operation above threshold, which suggests
$\Jmax\propto\left(1-\drrfrac\right)$, where $\Jmax$ is the current
flowing at $1'-3$ resonance ($\Delta_{1'3}=0$) and equation~\eqref{eq:drr2}
is used for substitution. For this IP QCL also, gain bandwidth broadening
with temperature is less likely to be the dominant temperature degradation
mechanism. Otherwise, the value of $\drrfrac$ should increase according to
equation~\eqref{eq:drr2} with the requirement of a larger population inversion
to reach threshold.

\begin{figure}[htbp]
\centering
\includegraphics[width=2.3in]{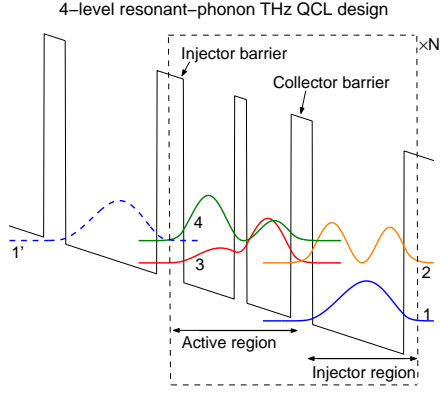}
\caption{
The 4-level resonant-phonon terahertz QCL structure reproduced from Fig.~\ref{fig:RPCBIVs}(a)
with a different level numbering scheme for simplicity of presentation in the following
discussions. The radiative transition
is from $4\rightarrow3$, and depopulation of the lower level is via $3\rightarrow2$ RT and
$2\rightarrow1$ e-LO phonon scattering where $E_{21}\approx\hwLO$.
}
\label{fig:4levCB}
\end{figure}

\section{\label{sec:4lev} Optical gain spectrum of resonant-phonon terahertz QCLs}

Terahertz QCL designs with RP depopulation typically have broad gain
linewidths. The spontaneous emission spectra from one of the earliest such designs had
a \mbox{$\sim2~\thz$} full-width half-maximum (FWHM) linewidth for a \mbox{$\sim4~\thz$
QCL active region~\cite{williams:thesis}}.
Experimental spontaneous emission data for some of the more recent high performance
designs~\cite{williams:copper,kumar:diagonal} is not available because of the
difficulty in observing sub-threshold optical signal in metal-metal
waveguides~\cite{williams:metal}, both because of the low loss in these
waveguides that causes lasing soon after the upper level is populated with
electrons, and also due to the poor out-coupling efficiency of such
cavities~\cite{kohen:waveguide}. However, simultaneous lasing of modes
separated by frequencies greater than $0.5~\thz$~\cite{williams:corrdfb,kumar:surfemit}
suggests that large gain exists over a broad bandwidth that is a significant
fraction of the center frequency in a homogeneously grown RPTQCL design
(in contrast to heterogeneous cascade designs that have been used to realize
broad gain bandwidths at mid-infrared frequencies~\cite{maulini:tunable}).

In this section, we use density-matrices to numerically estimate the optical
gain spectrum in a RPTQCL design and find reasons for broadening of the
gain spectrum. Similar to the \IV\ calculations
in section~\ref{sec:3lev}, the simplified DM calculations are shown to be an
effective method to understand the role of various design parameters toward the
optical response of a terahertz QCL gain medium, as opposed to the nonequilibrium
Green's function technique that incorporates a full quantum theory of gain~\cite{lee:neg1,wacker:gain}.
Similar calculations have been used recently to determine the optimal injector
coupling in mid-infrared QCLs~\cite{khurgin:iroughness}, and also to describe
the electrical transport characteristics of RT extraction based terahertz QCL
structure~\cite{scalari:tunnel}. We consider the simplest RP
design with 4-levels per QCL period as shown in Fig.~\ref{fig:4levCB}. To compute the
gain (or loss) spectrum, the linear response to a sinusoidal electric field perturbation
of the form $\bE = \zhat \, \mathcal{E} (\epiwt + \emiwt)$ is calculated by incorporating
the electric-dipole interaction term for coherent coupling of the radiative levels.
Only the electric-field component in the growth direction $\zhat$ is considered due
to the intersubband polarization selection rule. The time evolution of the DM for
the 4-level design is then written as~\cite{kumar:thesis1}

\scriptsize
\begin{widetext}
\begin{eqnarray}
\eightnegspc
\ddt
\begin{pmatrix}
\rho_{1'1'} & \rho_{1'4} & \rho_{1'3}\epiwt & \rho_{1'2}\epiwt \\
\rho_{41'} & \rho_{44} & \rho_{43}\epiwt & \rho_{42}\epiwt \\
\rho_{31'}\emiwt & \rho_{34}\emiwt & \rho_{33} & \rho_{32} \\
\rho_{21'}\emiwt & \rho_{24}\emiwt & \rho_{23} & \rho_{22} \\
\end{pmatrix}
\! \!
& = &
\! \!
-\frac{i}{\hbar}
\left[
\begin{pmatrix}
E_{1'} & -\hbar\Omega_{1'4} & 0 & 0\\
-\hbar\Omega_{1'4} & E_4 & |e| z_{43} \Ef \epiwt & 0 \\
0 & |e| z_{43} \Ef \emiwt & E_3 & -\hbar\Omega_{32} \\
0 & 0 & -\hbar\Omega_{32} & E_2 \\
\end{pmatrix}
,
\rhomat_{(1',4,3,2)}
\right]
\fournegspc
\nonumber \\ & &
\!
+
\begin{pmatrix}
\frac{\rho_{44}}{\tau_{41}} + \frac{\rho_{22}}{\tau_{21}} & \frac{-\rho_{1'4}}{\tdgen{14}} &
\frac{-\rho_{1'3}\epiwt}{\tdgen{13}} & \frac{-\rho_{1'2}\epiwt}{\tdgen{12}}
\\
\frac{-\rho_{41'}}{\tdgen{14}} & -\frac{\rho_{44}}{\tau_4}-\frac{\rho_{44}-\rho_{33}}{\tst} &
\frac{-\rho_{43}\epiwt}{\tdgen{34}} & \frac{-\rho_{42}\epiwt}{\tdgen{24}}
\\
\frac{-\rho_{31'}\emiwt}{\tdgen{13}} & -\frac{\rho_{34}\emiwt}{\tdgen{34}} &
\frac{\rho_{44}}{\tau_{43}} + \frac{\rho_{44}-\rho_{33}}{\tst} & \frac{-\rho_{32}}{\tdgen{23}}
\\
\frac{-\rho_{21'}\emiwt}{\tdgen{12}} & -\frac{\rho_{24}\emiwt}{\tdgen{24}} &
\frac{-\rho_{23}}{\tdgen{23}} & -\frac{\rho_{22}}{\tau_{21}}
\\
\end{pmatrix}
\label{eq:dm4lev}
\end{eqnarray}
\end{widetext}
\normalsize

Equation~\eqref{eq:dm4lev} is written similarly as equation~\eqref{eq:dm3lev2}
with additional modifications due to the sinusoidal electric-field perturbation. 
We consider linear response within the rotating-wave approximation (\ie\ $|\omega-\omega_{43}|\ll\omega$)
in which case the affected coherences can be explicitly written to have a sinusoidal time
variation as shown with the slowly-varying amplitudes $\rho_{mn}$.
The radiative levels $3$ and $4$ are now coherently coupled through the
off-diagonal electric-dipole interaction term $|e|\br\cdot\bE$ that 
has an amplitude $|e|z_{43} \Ef$, where $z_{43}=\bra{4} \hat{z} \ket{3}$ is the
dipole-matrix element for the said tightbinding levels. Since $(1',4)$ and $(3,2)$
are coherently coupled due to non-zero $\Omega_{1'4}$ and $\Omega_{32}$ terms respectively,
the coherences corresponding to $(1',3)$, $(1',2)$, and $(4,2)$ acquire a time-harmonic
character due to the time-harmonic $(4,3)$ coherent coupling. The sinusoidal component of
the coherences is written explicitly in the ansatz $\rho_{(1',4,3,2)}$, in which case
$\frac{d\rho_{mn}}{dt}=0$ in the steady-state for the slowly varying amplitudes $\rho_{mn}$.
Note that $\rho_{(1',4,3,2)}$ on the right-side of equation~\eqref{eq:dm4lev} is the same
as the left-side matrix appearing with the time derivative.

We made some simplifying approximations in writing equation~\eqref{eq:dm4lev}. Similar to
before, for simplicity of calculations, we assume a unity injection efficiency
($\Omega_{1'3}\approx0$), also a unity collection efficiency ($\Omega_{42}\approx0$), and
neglect the backscattering terms $\tau_{12}$ and $\tau_{34}$. For non-zero $\Omega_{1'3}$
and/or $\Omega_{42}$, equation~\eqref{eq:dm4lev} will become more complicated since some of
the non-harmonic terms in the $\rhomat_{(1',4,3,2)}$ matrix will additionally acquire
time-harmonic character and similarly some of the time-harmonic coherences will additionally
acquire a constant value in the steady-state, thereby effectively increasing the number of
independent variables to be solved for the system (currently $16$ for the written equation).
Also, a parameter $\tau_{41}$ is included to incorporate the effect of any indirect parasitic
scattering channels from $4$ to $1$, since direct $4\rightarrow1$ transport is otherwise
only allowed via inter-sub-module tunneling across the collector barrier for the chosen set
of tightbinding basis functions. We can now solve this equation, which, in addition to
yielding values of current and level populations, also yields the electrical
polarization $\zhat \, \Pol(t)$ induced due to the external optical field $\bE$ as
\begin{eqnarray}
\eightnegspc
\Pol & \equiv &
\epsilon_0 \chi(\omega) \mathcal{E} \epiwt \, + \, \cc
\nonumber \\ & = &
\frac{-|e| \langle \hat{z} \rangle}{\Vac}
\nonumber \\ & = &
\frac{-|e|z_{43}}{\Vac}
\left\{ 
\left[  \rho_{43}(\omega) + \rho_{34}(-\omega) \right]
\epiwt
\, + \, \cc
\right\}
\end{eqnarray}
where $\Vac$ is the volume of the active region. The induced electrical susceptibility
$\chi(\omega)$ becomes
\begin{eqnarray}
\chi(\omega) & \equiv &
\chi'(\omega) + i\chi''(\omega)
\nonumber \\ & = &
\frac{-|e|z_{43}}{\Vac\epsilon_0\mathcal{E}}
\left[  \rho_{43}(\omega) + \rho_{34}(-\omega) \right]
\label{eq:susceptibility}
\end{eqnarray}
which is independent of the amplitude $\Ef$ for small values. The optical
gain coefficient $g(\omega)$ (in meter$^{-1}$) is related to the imaginary part of
the susceptibility $\chi''(\omega)$ as
\begin{eqnarray}
g(\omega)|_{(\chi',\chi'') \ll n^2_\mathrm{r}} =
\frac{\chi''(\omega)}{n_\mathrm{r}} 
\frac{\omega}{c}
\label{eq:gainsusceptibility}
\end{eqnarray}
where $n_\mathrm{r}$ is the refractive index of the medium. In the following, we show
results of the computed gain spectrum as a function of several different parameters of interest.

\begin{figure}[htbp]
\centering
\includegraphics[width=2.4in]{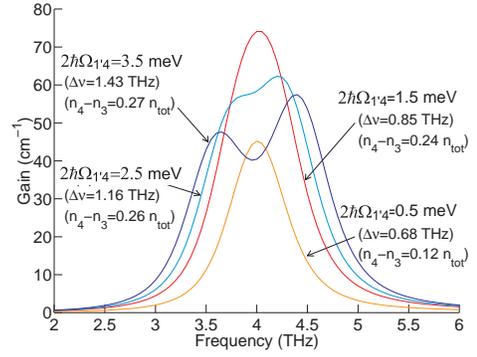}
\caption{
Computed gain spectra for the 4-level structure of Fig.~\ref{fig:4levCB} for different injector
anticrossings $2\hbar\Omega_{1'4}$, evaluated at $1'-4$ resonance ($\Delta_{1'4}=0$). Levels $3$
and $2$ are also taken to be at resonance ($\Delta_{32}=0$) and a small value of $2\hbar\Omega_{32}=2~\meV$
is chosen to limit additional broadening due to a coherent $3\rightarrow2$ RT process
(as shown in Fig.~\ref{fig:gainCA}a). All other parameters have same equivalent
values as in Fig.~\ref{fig:3levIVs}, except $\tau_{21}=0.2~\ps$ that results in an effective
lower state lifetime $\tau_{3,\rmeff}\left(\equiv \frac{n_3}{n_2}\tau_{21}\right)\approx1.0~\ps$,
which can also be estimated from the equivalent of equation~\eqref{eq:t2eff}.
Also, $E_{43}=16.5~\meV$ ($4~\thz$). The values of FWHM linewidth $\Delta\nu$, and
population inversion $\Delta n_{43}=n_4-n_3$ obtained as a result of the calculation are
also indicated alongside each of the curves.
}
\label{fig:gainIA}
\end{figure}

\subsection{\label{subsec:4levInj}Optical gain spectrum as a function of injector anticrossing}

Figure~\ref{fig:gainIA} shows the numerically computed gain spectra for the 4-level RPTQCL design
as modeled by equation~\eqref{eq:dm4lev} for different values of injector anticrossing
$2\hbar\Omega_{1'4}$. For the $4\rightarrow3$ radiative transition, the FWHM frequency linewidth
due to scattering is given by
$\Delta\nu_\mathrm{scatt} \approx
\frac{1}{\pi}
\left[
\frac{1}{2 \tau_{3,\mathrm{eff}}} +
\frac{1}{2 \tau_4} +
\frac{1}{T^*_2}
\right]$, which yields a value of $0.67~\thz$ ($h\Delta\nu_\mathrm{scatt}\sim2.75~\meV$)
for the chosen parameters. The value of $\Tpure=0.75~\ps$ was assumed for the above value
of the linewidth to approximately agree with those observed typically for bound-to-continuum
THz QCLs~\cite{scalari:btc,kroll:nature,jukam:gainbw} whose linewidths are more likely to be
scattering limited. As
$2\hbar\Omega_{1'4}$ becomes as large as $h\Delta\nu_\mathrm{scatt}$, injector transport becomes
more coherent and the gain spectrum becomes additionally broadened due to anticrossing
splitting of the linewidth as can be seen from Fig.~\ref{fig:gainIA}.
Also note that the population inversion $\Delta n$ approaches the maximum
value given by equation~\eqref{eq:dm3levCoh1} for coherent injection, which, for the
present case yields
$\Delta n_{\max}\sim\left(\frac{1-\tau_{3,\rmeff}/\tau_{43}}{2+\tau_{3,\rmeff}/\tau_{43}}\right)\;\ntot\approx0.28\;\ntot$.
The area under the gain
curve is proportional to the population inversion $\Delta n_{43}$, therefore, any additional
broadening decreases the peak gain. This suggests that a value of $2\hbar\Omega_{1'4}$ much larger
than $h\Delta\nu_\mathrm{scatt}$ might in fact diminish the performance of a design. However,
as will be shown in the next section,
RPTQCL designs anyway have broad linewidths due to a coherent RT assisted depopulation process,
in which case having $2\hbar\Omega_{1'4}>h\Delta\nu_\mathrm{scatt}$ is less likely to cause additional
broadening of the gain linewidth. Consequently, the injector anticrossing value for the RPTQCL
designs is chosen based on other design parameters such as the $1'\rightarrow3$ parasitic current
leakage~\cite{kumar:diagonal} rather than the concerns about gain broadening.

\begin{figure}[htbp]
\centering
\includegraphics[width=3.4in]{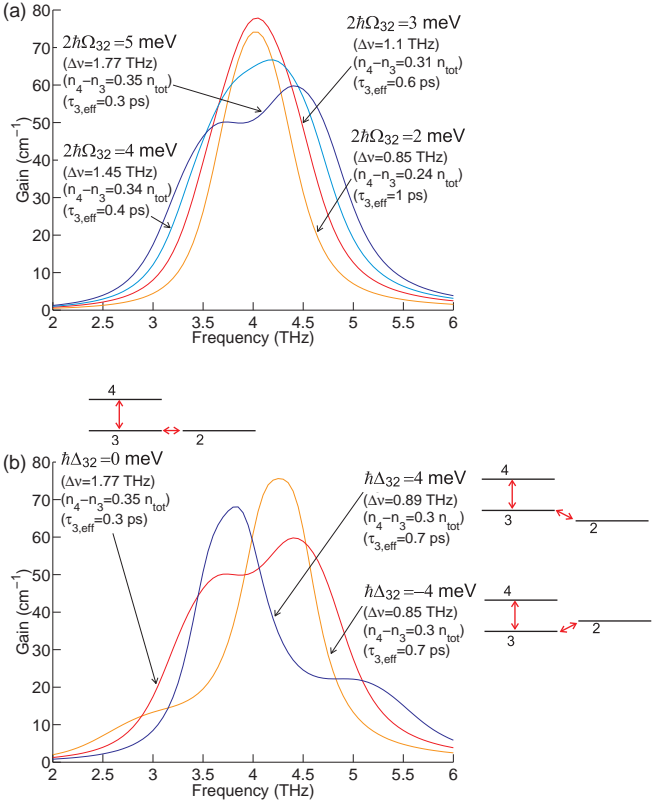}
\caption{
Computed gain spectra for the 4-level QCL design of Fig.~\ref{fig:4levCB} for (a) different values of
collector anticrossings $2\hbar\Omega_{32}$ at resonance ($\Delta_{32}=0$), and for (b) different values
of detuning $\hbar\Delta_{32}$ at $2\hbar\Omega_{32}=5~\meV$. Levels $1'$ and $4$ are assumed to be at resonance
($\Delta_{1'4}=0$) and a small value of $2\hbar\Omega_{1'4}=1.5~\meV$ is chosen to limit additional broadening
due to a coherent $1'\rightarrow4$ RT process (as shown in Fig.~\ref{fig:gainIA}). Other parameters
are the same as in Fig.~\ref{fig:gainIA}.  The values of FWHM linewidth $\Delta\nu$, effective
lifetime of the lower state $\tau_{3,\rmeff}\left(\equiv \frac{n_3}{n_2}\tau_{21}\right)$,
and population inversion $\Delta n_{43}=n_4-n_3$ obtained as a result of the calculation are
also indicated alongside each of the curves.
}
\label{fig:gainCA}
\end{figure}

\subsection{\label{subsec:4levColl}Optical gain spectrum as a function of collector anticrossing}

Figures~\ref{fig:gainCA}(a) and \ref{fig:gainCA}(b) show the computed gain spectra for different
values of collector anticrossings $2\hbar\Omega_{32}$, and energy detuning $\hbar\Delta_{32}$, respectively.
The best RPTQCL designs typically have collector anticrossing values in the range of $4-5~\meV$.
This is to maintain a short effective lower state lifetime so as to maximize $\Jmax/\Jth$, since
$\Jmax\propto\tau^{-1}_{\mathrm{\mathrm{lower}}}$ for coherent injection (equation~\ref{eq:dm3levCoh2}),
although at the cost of a broader linewidth.
For example, the $2\hbar\Omega_{32}=5~\meV$ spectrum in Fig.~\ref{fig:gainCA}(a) has a FWHM linewidth of
$~1.8~\thz$, a value almost twice as large as the scattering linewidth
$\Delta\nu_\mathrm{scatt}\sim 1~\thz$ (calculated using $\tau_{3,\rmeff}\sim0.3~\ps$).
This shows that the anticrossing splitting of the gain spectrum due to the coherent RT assisted
depopulation process is the main cause of the broad linewidths associated with
RPTQCLs~\cite{williams:thesis,williams:corrdfb,kumar:surfemit}.
As seen from Fig.~\ref{fig:gainCA}(b), note that linewidth broadening happens only close
to the $3-2$ resonance at $\hbar\Delta_{32}\approx0$. For $|\hbar\Delta_{32}|> 0$ and $|\hbar\Delta_{32}|\not\ll E_{43}$,
the width of the spectrum remains narrow, since the $4\rightarrow2$ radiative transition (which is
indirect, and due to $4\leftrightarrow3$ and $3\leftrightarrow2$ coherent coupling) is considerably
detuned from the stronger $4\rightarrow3$ radiative transition (which is direct, and due to
$4\leftrightarrow3$ coherent coupling), and hence contributes negligibly to the gain.
For a coherent $3\rightarrow2$ RT process (\ie\ large $\Omega_{32}$), the peak gain
occurs at a frequency $\omega_\rmpeak > E_{43}/\hbar$ for $\hbar\Delta_{32}<0$ and $\omega_\rmpeak < E_{43}/\hbar$
for $\hbar\Delta_{32}>0$. The frequency shift of $\omega_\rmpeak$ away from $E_{43}/\hbar$ is due to the    
energy splitting due to the $3-2$ anticrossing close to resonance. 

\begin{figure}[htbp]
\centering
\includegraphics[width=3.0in]{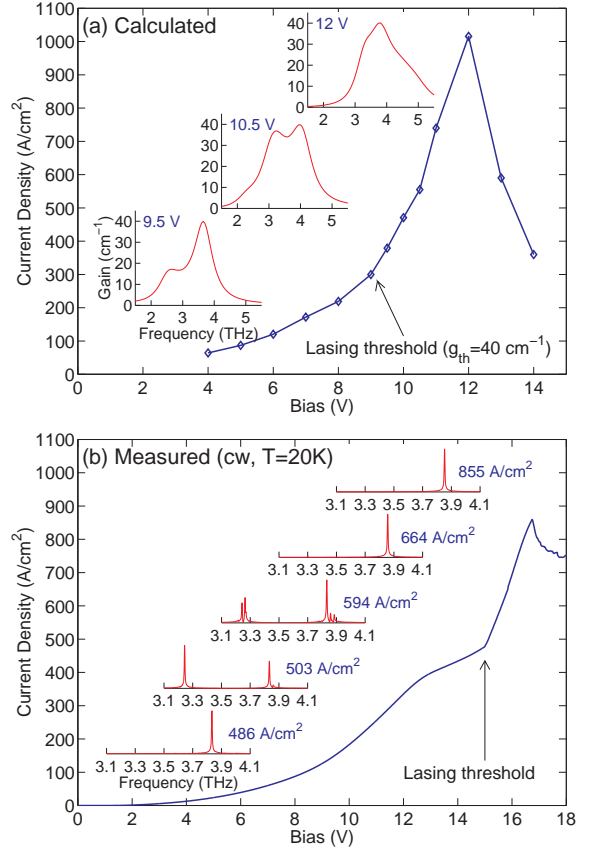}
\caption{(a) Computed \IV, and gain spectra at selected bias points for the 4-level QCL
in Ref.~\onlinecite{kumar:diagonal}. These results are obtained by solving
equation~\eqref{eq:dm4lev} in the steady-state $\left(\frac{d}{dt} \rightarrow 0\right)$, whereby,
$I\equiv|e|\left( \frac{\rho_{44}}{\tau_{41}} + \frac{\rho_{22}}{\tau_{21}} \right)$ and the
gain is calculated using equation~\eqref{eq:gainsusceptibility}. A value
of $\gth=40~\icm$ is assumed similar to that for Fig.~\ref{fig:3levIVs}~\cite{RTpaperf1:footnote},
following which the stimulated lifetime $\tst$ is determined to keep the peak gain constant as $\gth$
beyond threshold. Non-radiative lifetimes are calculated by including e-LO phonon
scattering only whereby an electron temperature of $100~\rmK$ is assumed in the upper level~$4$.
Also, $\Tpure\sim0.75~\ps$, and $\Omega_{1'3}\approx\Omega_{42}\approx 0$. (b) Experimental
\IV, and spectra (shown as insets) at selected bias points measured at a heat-sink temperature
of $20~\rmK$ in cw mode. The device is a $100~\um\times 1.39~\rmmm$ metal-metal waveguide ridge
laser from the same active region as in Ref.~\onlinecite{kumar:diagonal}. The higher than expected
bias voltages for the experimental \IV\ are likely due to additional voltage drop at the
electrical contacts.
}
\label{fig:gainOWI222G}
\end{figure}

We now show computed gain spectra of the 4-level RPTQCL design from
Ref.~\onlinecite{kumar:diagonal} in Fig.~\ref{fig:gainOWI222G}(a) as a function of
applied electrical bias. For a comparison, experimentally measured cw \IV\ and lasing
spectra of a ridge laser processed from the same active medium is also shown in
Fig.~\ref{fig:gainOWI222G}(b). For the calculation in Fig.~\ref{fig:gainOWI222G}(a),
the radiative energy $E_{43}$ varies linearly from $12.4~\meV$ ($3.0~\thz$)
at $8~\rmV$ to $15.7~\meV$ ($3.8~\thz$) at the $1'-4$ resonance bias of $12~\rmV$ due
to the large value of the Stark shift in a diagonal radiative transition. However, the
peak gain occurs close to a frequency of $4~\thz$ at almost all bias conditions for
the calculated spectra. This behavior is due to the coherent nature of the
depopulation process. The $3-2$ energy detuning $\hbar\Delta_{32}$ also varies linearly
with the applied bias, from $-4.7~\meV$ at $8~\rmV$ to $2.4~\meV$ at $12~\rmV$.
Consequently, as shown by the calculations in Fig.~\ref{fig:gainCA}(b), the gain
peak is pushed to higher frequencies than the radiative separation $E_{43}$ at
low-bias that effectively masks the Stark shift in $E_{43}$. This is affirmed
from the lasing spectra in Fig.~\ref{fig:gainOWI222G}(b) that shows that the
$\sim3.9~\thz$ mode is excited at all bias. In contrast, the Stark shift
is clearly emphasized in the lasing mode spectrum of the diagonal
bound-to-continuum terahertz QCLs~\cite{scalari:review} due to the
different depopulation mechanism in such designs.

As an evidence of the broad gain linewidth in this QCL, at some middle-bias
points, additional lasing spectra are excited close to a frequency of $3.3~\thz$
that leads to simultaneous cw lasing of modes separated by $\sim0.6~\thz$.
This behavior could be attributed
to the double-peaked gain spectra that emerges close to the $3-2$ resonance
($\Delta_{32}\approx0$), which, for this design, happens at the middle-bias points.
It may be noted that we have observed similar dual-frequency lasing behavior in
some versions of the 5-level RPTQCL design of Fig.~\ref{fig:RPCBIVs}(b). Whether
or not this happens depends on the relative alignment of the injector and
collector anticrossings as a function of the applied electrical bias.

We should note that recently a much smaller FWHM linewidth of $\sim0.6~\thz$ has been
measured for a similar 4-level RPTQCL design~\cite{jukam:tdsphononqcl}
using a THz time-domain spectroscopy technique, which could be due to the 
following reasons. The QCL structure in Ref.~\onlinecite{jukam:tdsphononqcl} has a
smaller collector anticrossing $2\hbar\Omega_{32}\sim3.7~\meV$ as opposed to a value of
$2\hbar\Omega_{32}\sim4.7~\meV$ for the design in Ref.~\onlinecite{kumar:diagonal} that is
analyzed in Fig.~\ref{fig:gainOWI222G}. Also, it is possible for a given QCL structure
that the bias range of lasing operation may not sweep through its collector resonance
($\Delta_{32}=0$) depending on the relative alignment of the injector and the collector
anticrossings for the grown structure, in which case
the gain spectra may not show additional broadening due to the $3\rightarrow2$ RT as
seen from some specific calculations for non-zero $\hbar\Delta_{32}$ in Fig.~\ref{fig:gainCA}(b).
We would also like to note that the present calculations were done within a
rotating-wave approximation ($|\omega-\omega_{43}|\ll\omega$), which becomes less accurate
once the gain linewidth
becomes a significant fraction of the center frequency. Hence, the true linewidths
may be somewhat narrower than those calculated in this section.
Nevertheless, within the assumptions considered, the close tracking of the calculated
and experimental spectral characteristics of the 4-level RPTQCL design in
Fig.~\ref{fig:gainOWI222G} establishes the importance of the relatively simple
density-matrix model developed in this section for estimating the optical gain
spectrum of a terahertz QCL structure.

In conclusion, we have presented simplified density-matrix transport models to
describe resonant-tunneling transport in terahertz QCLs. Due to the closely
spaced energy levels in terahertz QCL structures, coherence plays an important
role in the resonant-tunneling mechanism, which is incorporated well within the
presented framework. A 3-level model was developed to derive current transport
through the injector barrier of any general QCL design (which applies to
mid-infrared QCLs as well). Useful expressions were derived for current flow
above and below threshold that could directly be used to analyze the experimental
behavior of some representative QCL designs. Based on experimental observations,
we have been able to speculate about some of the dominant temperature degradation
mechanisms in phonon-depopulated terahertz QCL designs. We have extended the 
density-matrix model to estimate the gain spectra of resonant-phonon terahertz
QCLs. A coherent resonant-tunneling assisted depopulation process is
identified to be the primary cause of the broad gain bandwidths typically
observed in such QCLs.

\begin{acknowledgments}
We would like to thank Chun W. I. Chan for assistance with some of the experimental
work reported here. This work is supported by AFOSR, NASA, and NSF.
\end{acknowledgments}



\begin{thebibliography}{46}
\expandafter\ifx\csname natexlab\endcsname\relax\def\natexlab#1{#1}\fi
\expandafter\ifx\csname bibnamefont\endcsname\relax
  \def\bibnamefont#1{#1}\fi
\expandafter\ifx\csname bibfnamefont\endcsname\relax
  \def\bibfnamefont#1{#1}\fi
\expandafter\ifx\csname citenamefont\endcsname\relax
  \def\citenamefont#1{#1}\fi
\expandafter\ifx\csname url\endcsname\relax
  \def\url#1{\texttt{#1}}\fi
\expandafter\ifx\csname urlprefix\endcsname\relax\def\urlprefix{URL }\fi
\providecommand{\bibinfo}[2]{#2}
\providecommand{\eprint}[2][]{\url{#2}}

\bibitem[{\citenamefont{Esaki and Tsu}(1970)}]{esaki:sl}
\bibinfo{author}{\bibfnamefont{L.}~\bibnamefont{Esaki}} \bibnamefont{and}
  \bibinfo{author}{\bibfnamefont{R.}~\bibnamefont{Tsu}}, \bibinfo{journal}{IBM
  J. Res. Dev.} \textbf{\bibinfo{volume}{14}}, \bibinfo{pages}{61}
  (\bibinfo{year}{1970}).

\bibitem[{\citenamefont{Kazarinov and Suris}(1971)}]{kazarinov:sl1}
\bibinfo{author}{\bibfnamefont{R.~F.} \bibnamefont{Kazarinov}}
  \bibnamefont{and} \bibinfo{author}{\bibfnamefont{R.~A.} \bibnamefont{Suris}},
  \bibinfo{journal}{Sov. Phys. Semicond.} \textbf{\bibinfo{volume}{5}},
  \bibinfo{pages}{707} (\bibinfo{year}{1971}).

\bibitem[{\citenamefont{Faist et~al.}(1994)\citenamefont{Faist, Capasso, Sivco,
  Sirtori, Hutchinson, and Cho}}]{faist:qcl}
\bibinfo{author}{\bibfnamefont{J.}~\bibnamefont{Faist}},
  \bibinfo{author}{\bibfnamefont{F.}~\bibnamefont{Capasso}},
  \bibinfo{author}{\bibfnamefont{D.~L.} \bibnamefont{Sivco}},
  \bibinfo{author}{\bibfnamefont{C.}~\bibnamefont{Sirtori}},
  \bibinfo{author}{\bibfnamefont{A.~L.} \bibnamefont{Hutchinson}},
  \bibnamefont{and} \bibinfo{author}{\bibfnamefont{A.~Y.} \bibnamefont{Cho}},
  \bibinfo{journal}{Science} \textbf{\bibinfo{volume}{264}},
  \bibinfo{pages}{553} (\bibinfo{year}{1994}).

\bibitem[{\citenamefont{K\"{o}hler et~al.}(2002)\citenamefont{K\"{o}hler,
  Tredicucci, Beltram, Beere, Linfield, Davies, Ritchie, Iotti, and
  Rossi}}]{kohler:laser}
\bibinfo{author}{\bibfnamefont{R.}~\bibnamefont{K\"{o}hler}},
  \bibinfo{author}{\bibfnamefont{A.}~\bibnamefont{Tredicucci}},
  \bibinfo{author}{\bibfnamefont{F.}~\bibnamefont{Beltram}},
  \bibinfo{author}{\bibfnamefont{H.~E.} \bibnamefont{Beere}},
  \bibinfo{author}{\bibfnamefont{E.~H.} \bibnamefont{Linfield}},
  \bibinfo{author}{\bibfnamefont{A.~G.} \bibnamefont{Davies}},
  \bibinfo{author}{\bibfnamefont{D.~A.} \bibnamefont{Ritchie}},
  \bibinfo{author}{\bibfnamefont{R.~C.} \bibnamefont{Iotti}}, \bibnamefont{and}
  \bibinfo{author}{\bibfnamefont{F.}~\bibnamefont{Rossi}},
  \bibinfo{journal}{Nature} \textbf{\bibinfo{volume}{417}},
  \bibinfo{pages}{156} (\bibinfo{year}{2002}).

\bibitem[{\citenamefont{Williams}(2007)}]{williams:review}
\bibinfo{author}{\bibfnamefont{B.~S.} \bibnamefont{Williams}},
  \bibinfo{journal}{Nature Photonics} \textbf{\bibinfo{volume}{1}},
  \bibinfo{pages}{517} (\bibinfo{year}{2007}).

\bibitem[{\citenamefont{Kumar and Lee}(2008)}]{kumar:review08}
\bibinfo{author}{\bibfnamefont{S.}~\bibnamefont{Kumar}} \bibnamefont{and}
  \bibinfo{author}{\bibfnamefont{A.~W.~M.} \bibnamefont{Lee}},
  \bibinfo{journal}{IEEE J. Sel. Topics Quantum Electron.}
  \textbf{\bibinfo{volume}{14}}, \bibinfo{pages}{333} (\bibinfo{year}{2008}).

\bibitem[{\citenamefont{Scalari
  et~al.}(2009{\natexlab{a}})\citenamefont{Scalari, Walther, Fischer, Terazzi,
  Beere, Ritchie, and Faist}}]{scalari:review}
\bibinfo{author}{\bibfnamefont{G.}~\bibnamefont{Scalari}},
  \bibinfo{author}{\bibfnamefont{C.}~\bibnamefont{Walther}},
  \bibinfo{author}{\bibfnamefont{M.}~\bibnamefont{Fischer}},
  \bibinfo{author}{\bibfnamefont{R.}~\bibnamefont{Terazzi}},
  \bibinfo{author}{\bibfnamefont{H.}~\bibnamefont{Beere}},
  \bibinfo{author}{\bibfnamefont{D.}~\bibnamefont{Ritchie}}, \bibnamefont{and}
  \bibinfo{author}{\bibfnamefont{J.}~\bibnamefont{Faist}},
  \bibinfo{journal}{Laser Photonics Rev.} \textbf{\bibinfo{volume}{3}},
  \bibinfo{pages}{45} (\bibinfo{year}{2009}{\natexlab{a}}).

\bibitem[{\citenamefont{Williams
  et~al.}(2003{\natexlab{a}})\citenamefont{Williams, Callebaut, Kumar, Hu, and
  Reno}}]{williams:laser}
\bibinfo{author}{\bibfnamefont{B.~S.} \bibnamefont{Williams}},
  \bibinfo{author}{\bibfnamefont{H.}~\bibnamefont{Callebaut}},
  \bibinfo{author}{\bibfnamefont{S.}~\bibnamefont{Kumar}},
  \bibinfo{author}{\bibfnamefont{Q.}~\bibnamefont{Hu}}, \bibnamefont{and}
  \bibinfo{author}{\bibfnamefont{J.~L.} \bibnamefont{Reno}},
  \bibinfo{journal}{Appl. Phys. Lett.} \textbf{\bibinfo{volume}{82}},
  \bibinfo{pages}{1015} (\bibinfo{year}{2003}{\natexlab{a}}).

\bibitem[{\citenamefont{Hu et~al.}(2005)\citenamefont{Hu, Williams, Kumar,
  Callebaut, Kohen, and Reno}}]{hu:respho}
\bibinfo{author}{\bibfnamefont{Q.}~\bibnamefont{Hu}},
  \bibinfo{author}{\bibfnamefont{B.~S.} \bibnamefont{Williams}},
  \bibinfo{author}{\bibfnamefont{S.}~\bibnamefont{Kumar}},
  \bibinfo{author}{\bibfnamefont{H.}~\bibnamefont{Callebaut}},
  \bibinfo{author}{\bibfnamefont{S.}~\bibnamefont{Kohen}}, \bibnamefont{and}
  \bibinfo{author}{\bibfnamefont{J.~L.} \bibnamefont{Reno}},
  \bibinfo{journal}{Semicond. Sci. Technol.} \textbf{\bibinfo{volume}{20}},
  \bibinfo{pages}{S228} (\bibinfo{year}{2005}).

\bibitem[{\citenamefont{Kumar et~al.}(2009{\natexlab{a}})\citenamefont{Kumar,
  Hu, and Reno}}]{kumar:diagonal}
\bibinfo{author}{\bibfnamefont{S.}~\bibnamefont{Kumar}},
  \bibinfo{author}{\bibfnamefont{Q.}~\bibnamefont{Hu}}, \bibnamefont{and}
  \bibinfo{author}{\bibfnamefont{J.~L.} \bibnamefont{Reno}},
  \bibinfo{journal}{Appl. Phys. Lett.} \textbf{\bibinfo{volume}{94}},
  \bibinfo{pages}{131105} (\bibinfo{year}{2009}{\natexlab{a}}).

\bibitem[{\citenamefont{Wade et~al.}(2009)\citenamefont{Wade, Fedorov, Smirnov,
  Kumar, Williams, Hu, and Reno}}]{wade:thzqcl}
\bibinfo{author}{\bibfnamefont{A.}~\bibnamefont{Wade}},
  \bibinfo{author}{\bibfnamefont{G.}~\bibnamefont{Fedorov}},
  \bibinfo{author}{\bibfnamefont{D.}~\bibnamefont{Smirnov}},
  \bibinfo{author}{\bibfnamefont{S.}~\bibnamefont{Kumar}},
  \bibinfo{author}{\bibfnamefont{B.~S.} \bibnamefont{Williams}},
  \bibinfo{author}{\bibfnamefont{Q.}~\bibnamefont{Hu}}, \bibnamefont{and}
  \bibinfo{author}{\bibfnamefont{J.~L.} \bibnamefont{Reno}},
  \bibinfo{journal}{Nature Photonics} \textbf{\bibinfo{volume}{3}},
  \bibinfo{pages}{41} (\bibinfo{year}{2009}).

\bibitem[{\citenamefont{Sirtori et~al.}(1998)\citenamefont{Sirtori, Capasso,
  Faist, Hutchinson, Sivco, and Cho}}]{sirtori:tunnel}
\bibinfo{author}{\bibfnamefont{C.}~\bibnamefont{Sirtori}},
  \bibinfo{author}{\bibfnamefont{F.}~\bibnamefont{Capasso}},
  \bibinfo{author}{\bibfnamefont{J.}~\bibnamefont{Faist}},
  \bibinfo{author}{\bibfnamefont{A.~L.} \bibnamefont{Hutchinson}},
  \bibinfo{author}{\bibfnamefont{D.~L.} \bibnamefont{Sivco}}, \bibnamefont{and}
  \bibinfo{author}{\bibfnamefont{A.~Y.} \bibnamefont{Cho}},
  \bibinfo{journal}{IEEE J. Quantum Electron.} \textbf{\bibinfo{volume}{34}},
  \bibinfo{pages}{1722} (\bibinfo{year}{1998}).

\bibitem[{\citenamefont{Terazzi et~al.}(2008)\citenamefont{Terazzi, Gresch,
  Wittmann, and Faist}}]{terazzi:tunnel}
\bibinfo{author}{\bibfnamefont{R.}~\bibnamefont{Terazzi}},
  \bibinfo{author}{\bibfnamefont{T.}~\bibnamefont{Gresch}},
  \bibinfo{author}{\bibfnamefont{A.}~\bibnamefont{Wittmann}}, \bibnamefont{and}
  \bibinfo{author}{\bibfnamefont{J.}~\bibnamefont{Faist}},
  \bibinfo{journal}{Phys. Rev. B} \textbf{\bibinfo{volume}{78}},
  \bibinfo{pages}{155328} (\bibinfo{year}{2008}).

\bibitem[{\citenamefont{Scalari et~al.}(2007)\citenamefont{Scalari, Terazzi,
  Giovannini, Hoyler, and Faist}}]{scalari:tunnel}
\bibinfo{author}{\bibfnamefont{G.}~\bibnamefont{Scalari}},
  \bibinfo{author}{\bibfnamefont{R.}~\bibnamefont{Terazzi}},
  \bibinfo{author}{\bibfnamefont{M.}~\bibnamefont{Giovannini}},
  \bibinfo{author}{\bibfnamefont{N.}~\bibnamefont{Hoyler}}, \bibnamefont{and}
  \bibinfo{author}{\bibfnamefont{J.}~\bibnamefont{Faist}},
  \bibinfo{journal}{Appl. Phys. Lett.} \textbf{\bibinfo{volume}{91}},
  \bibinfo{pages}{032103} (\bibinfo{year}{2007}).

\bibitem[{\citenamefont{Scalari
  et~al.}(2009{\natexlab{b}})\citenamefont{Scalari, Amanti, Fischer, Terazzi,
  Walther, Beck, and Faist}}]{scalari:stepwell}
\bibinfo{author}{\bibfnamefont{G.}~\bibnamefont{Scalari}},
  \bibinfo{author}{\bibfnamefont{M.~I.} \bibnamefont{Amanti}},
  \bibinfo{author}{\bibfnamefont{M.}~\bibnamefont{Fischer}},
  \bibinfo{author}{\bibfnamefont{R.}~\bibnamefont{Terazzi}},
  \bibinfo{author}{\bibfnamefont{C.}~\bibnamefont{Walther}},
  \bibinfo{author}{\bibfnamefont{M.}~\bibnamefont{Beck}}, \bibnamefont{and}
  \bibinfo{author}{\bibfnamefont{J.}~\bibnamefont{Faist}},
  \bibinfo{journal}{Appl. Phys. Lett.} \textbf{\bibinfo{volume}{94}},
  \bibinfo{pages}{041114} (\bibinfo{year}{2009}{\natexlab{b}}).

\bibitem[{\citenamefont{Callebaut and Hu}(2005)}]{callebaut:mcdm}
\bibinfo{author}{\bibfnamefont{H.}~\bibnamefont{Callebaut}} \bibnamefont{and}
  \bibinfo{author}{\bibfnamefont{Q.}~\bibnamefont{Hu}}, \bibinfo{journal}{J.
  Appl. Phys.} \textbf{\bibinfo{volume}{98}}, \bibinfo{pages}{104505}
  (\bibinfo{year}{2005}).

\bibitem[{\citenamefont{Kazarinov and Suris}(1972)}]{kazarinov:sl2}
\bibinfo{author}{\bibfnamefont{R.~F.} \bibnamefont{Kazarinov}}
  \bibnamefont{and} \bibinfo{author}{\bibfnamefont{R.~A.} \bibnamefont{Suris}},
  \bibinfo{journal}{Sov. Phys. Semicond.} \textbf{\bibinfo{volume}{6}},
  \bibinfo{pages}{120} (\bibinfo{year}{1972}).

\bibitem[{\citenamefont{Willenberg et~al.}(2003)\citenamefont{Willenberg,
  D\"{o}hler, and Faist}}]{willenberg:bloch}
\bibinfo{author}{\bibfnamefont{H.}~\bibnamefont{Willenberg}},
  \bibinfo{author}{\bibfnamefont{G.~H.} \bibnamefont{D\"{o}hler}},
  \bibnamefont{and} \bibinfo{author}{\bibfnamefont{J.}~\bibnamefont{Faist}},
  \bibinfo{journal}{Phys. Rev. B} \textbf{\bibinfo{volume}{67}},
  \bibinfo{pages}{085315} (\bibinfo{year}{2003}).

\bibitem[{\citenamefont{Williams}(2003)}]{williams:thesis}
\bibinfo{author}{\bibfnamefont{B.~S.} \bibnamefont{Williams}},
  \bibinfo{type}{{PhD} dissertation}, \bibinfo{school}{Massachusetts Institute
  of Technology}, \bibinfo{address}{Department of Electrical Engineering and
  Computer Science} (\bibinfo{year}{2003}).

\bibitem[{\citenamefont{Kumar et~al.}(2007)\citenamefont{Kumar, Williams, Qin,
  Lee, Hu, and Reno}}]{kumar:surfemit}
\bibinfo{author}{\bibfnamefont{S.}~\bibnamefont{Kumar}},
  \bibinfo{author}{\bibfnamefont{B.~S.} \bibnamefont{Williams}},
  \bibinfo{author}{\bibfnamefont{Q.}~\bibnamefont{Qin}},
  \bibinfo{author}{\bibfnamefont{A.~W.~M.} \bibnamefont{Lee}},
  \bibinfo{author}{\bibfnamefont{Q.}~\bibnamefont{Hu}}, \bibnamefont{and}
  \bibinfo{author}{\bibfnamefont{J.~L.} \bibnamefont{Reno}},
  \bibinfo{journal}{Opt. Express} \textbf{\bibinfo{volume}{15}},
  \bibinfo{pages}{113} (\bibinfo{year}{2007}).

\bibitem[{\citenamefont{Lee et~al.}(2006)\citenamefont{Lee, Banit, Woerner, and
  Wacker}}]{lee:neg2}
\bibinfo{author}{\bibfnamefont{S.-C.} \bibnamefont{Lee}},
  \bibinfo{author}{\bibfnamefont{F.}~\bibnamefont{Banit}},
  \bibinfo{author}{\bibfnamefont{M.}~\bibnamefont{Woerner}}, \bibnamefont{and}
  \bibinfo{author}{\bibfnamefont{A.}~\bibnamefont{Wacker}},
  \bibinfo{journal}{Phys. Rev. B} \textbf{\bibinfo{volume}{73}},
  \bibinfo{pages}{245320} (\bibinfo{year}{2006}).

\bibitem[{\citenamefont{Kumar}(2007)}]{kumar:thesis1}
\bibinfo{author}{\bibfnamefont{S.}~\bibnamefont{Kumar}}, \bibinfo{type}{{PhD}
  dissertation}, \bibinfo{school}{Massachusetts Institute of Technology},
  \bibinfo{address}{Department of Electrical Engineering and Computer Science}
  (\bibinfo{year}{2007}),
  \bibinfo{note}{(http://dspace.mit.edu/handle/1721.1/40501)}.

\bibitem[{\citenamefont{Kumar et~al.}(2009{\natexlab{b}})\citenamefont{Kumar,
  Chan, Hu, and Reno}}]{kumar:twowell}
\bibinfo{author}{\bibfnamefont{S.}~\bibnamefont{Kumar}},
  \bibinfo{author}{\bibfnamefont{C.~W.~I.} \bibnamefont{Chan}},
  \bibinfo{author}{\bibfnamefont{Q.}~\bibnamefont{Hu}}, \bibnamefont{and}
  \bibinfo{author}{\bibfnamefont{J.~L.} \bibnamefont{Reno}},
  \bibinfo{journal}{Appl. Phys. Lett.} \textbf{\bibinfo{volume}{95}},
  \bibinfo{pages}{141110} (\bibinfo{year}{2009}).

\bibitem[{\citenamefont{Eickemeyer et~al.}(2002)\citenamefont{Eickemeyer,
  Reimann, Woerner, Elsaesser, Barbieri, Sirtori, Strasser, M\"{u}ller,
  Bratschitsch, and Unterrainer}}]{eickemeyer:coherent}
\bibinfo{author}{\bibfnamefont{F.}~\bibnamefont{Eickemeyer}},
  \bibinfo{author}{\bibfnamefont{K.}~\bibnamefont{Reimann}},
  \bibinfo{author}{\bibfnamefont{M.}~\bibnamefont{Woerner}},
  \bibinfo{author}{\bibfnamefont{T.}~\bibnamefont{Elsaesser}},
  \bibinfo{author}{\bibfnamefont{S.}~\bibnamefont{Barbieri}},
  \bibinfo{author}{\bibfnamefont{C.}~\bibnamefont{Sirtori}},
  \bibinfo{author}{\bibfnamefont{G.}~\bibnamefont{Strasser}},
  \bibinfo{author}{\bibfnamefont{T.}~\bibnamefont{M\"{u}ller}},
  \bibinfo{author}{\bibfnamefont{R.}~\bibnamefont{Bratschitsch}},
  \bibnamefont{and}
  \bibinfo{author}{\bibfnamefont{K.}~\bibnamefont{Unterrainer}},
  \bibinfo{journal}{Phys. Rev. Lett.} \textbf{\bibinfo{volume}{89}},
  \bibinfo{pages}{047402} (\bibinfo{year}{2002}).

\bibitem[{RTp()}]{RTpaperf1:footnote}
\bibinfo{note}{The value of $40~\icm$ for the threshold material gain $\gth$
  was chosen based on some published estimates. For single-plasmon
  waveguides~\cite{kohler:laser}, values of $\sim25~\icm$ have been estimated
  experimentally for the modal gain threshold of $1~\rmmm$ long cavities with a
  $3.1~\thz$~\cite{jukam:tdsphononqcl} and a $4.3~\thz$~\cite{kumar:ptw08}
  resonant-phonon QCL gain medium, respectively. This results in $\gth>70~\icm$
  for such waveguides given their low mode-confinement factors
  ($25-35\%$~\cite{kohen:waveguide}). In contrast, metal-metal
  waveguides~\cite{williams:metal} have lower losses~\cite{kohen:waveguide},
  with the lone published experimental estimate being $\gth\sim36\pm10~\icm$
  for a $3.6~\thz$ bound-to-continuum QCL~\cite{dunbar:microcavity}.}

\bibitem[{\citenamefont{Luo et~al.}(2007)\citenamefont{Luo, Laframboise,
  Wasilewski, Aers, Liu, and Cao}}]{luo:threewell}
\bibinfo{author}{\bibfnamefont{H.}~\bibnamefont{Luo}},
  \bibinfo{author}{\bibfnamefont{S.~R.} \bibnamefont{Laframboise}},
  \bibinfo{author}{\bibfnamefont{Z.~R.} \bibnamefont{Wasilewski}},
  \bibinfo{author}{\bibfnamefont{G.~C.} \bibnamefont{Aers}},
  \bibinfo{author}{\bibfnamefont{H.~C.} \bibnamefont{Liu}}, \bibnamefont{and}
  \bibinfo{author}{\bibfnamefont{J.~C.} \bibnamefont{Cao}},
  \bibinfo{journal}{Appl. Phys. Lett.} \textbf{\bibinfo{volume}{90}},
  \bibinfo{pages}{041112} (\bibinfo{year}{2007}).

\bibitem[{\citenamefont{Kumar et~al.}(2006)\citenamefont{Kumar, Williams, , Hu,
  and Reno}}]{kumar:owi}
\bibinfo{author}{\bibfnamefont{S.}~\bibnamefont{Kumar}},
  \bibinfo{author}{\bibfnamefont{B.~S.} \bibnamefont{Williams}}, ,
  \bibinfo{author}{\bibfnamefont{Q.}~\bibnamefont{Hu}}, \bibnamefont{and}
  \bibinfo{author}{\bibfnamefont{J.~L.} \bibnamefont{Reno}},
  \bibinfo{journal}{Appl. Phys. Lett.} \textbf{\bibinfo{volume}{88}},
  \bibinfo{pages}{121123} (\bibinfo{year}{2006}).

\bibitem[{\citenamefont{Nelander and Wacker}(2008)}]{nelander:broadening}
\bibinfo{author}{\bibfnamefont{R.}~\bibnamefont{Nelander}} \bibnamefont{and}
  \bibinfo{author}{\bibfnamefont{A.}~\bibnamefont{Wacker}},
  \bibinfo{journal}{Appl. Phys. Lett.} \textbf{\bibinfo{volume}{92}},
  \bibinfo{pages}{081102} (\bibinfo{year}{2008}).

\bibitem[{\citenamefont{Ajili et~al.}(2004)\citenamefont{Ajili, Scalari, Faist,
  Beere, Linfield, Ritchie, and Davies}}]{ajili:highpower}
\bibinfo{author}{\bibfnamefont{L.}~\bibnamefont{Ajili}},
  \bibinfo{author}{\bibfnamefont{G.}~\bibnamefont{Scalari}},
  \bibinfo{author}{\bibfnamefont{J.}~\bibnamefont{Faist}},
  \bibinfo{author}{\bibfnamefont{H.}~\bibnamefont{Beere}},
  \bibinfo{author}{\bibfnamefont{E.}~\bibnamefont{Linfield}},
  \bibinfo{author}{\bibfnamefont{D.}~\bibnamefont{Ritchie}}, \bibnamefont{and}
  \bibinfo{author}{\bibfnamefont{G.}~\bibnamefont{Davies}},
  \bibinfo{journal}{Appl. Phys. Lett.} \textbf{\bibinfo{volume}{85}},
  \bibinfo{pages}{3986} (\bibinfo{year}{2004}).

\bibitem[{\citenamefont{Chang et~al.}(1974)\citenamefont{Chang, Esaki, and
  Tsu}}]{chang:rt}
\bibinfo{author}{\bibfnamefont{L.~L.} \bibnamefont{Chang}},
  \bibinfo{author}{\bibfnamefont{L.}~\bibnamefont{Esaki}}, \bibnamefont{and}
  \bibinfo{author}{\bibfnamefont{R.}~\bibnamefont{Tsu}},
  \bibinfo{journal}{Appl. Phys. Lett.} \textbf{\bibinfo{volume}{24}},
  \bibinfo{pages}{593} (\bibinfo{year}{1974}).

\bibitem[{\citenamefont{Ohno et~al.}(1990)\citenamefont{Ohno, Mendez, and
  Wang}}]{ohno:rtd}
\bibinfo{author}{\bibfnamefont{H.}~\bibnamefont{Ohno}},
  \bibinfo{author}{\bibfnamefont{E.~E.} \bibnamefont{Mendez}},
  \bibnamefont{and} \bibinfo{author}{\bibfnamefont{W.~I.} \bibnamefont{Wang}},
  \bibinfo{journal}{Appl. Phys. Lett.} \textbf{\bibinfo{volume}{56}},
  \bibinfo{pages}{1793} (\bibinfo{year}{1990}).

\bibitem[{\citenamefont{Wacker}(2001)}]{wacker:transport}
\bibinfo{author}{\bibfnamefont{A.}~\bibnamefont{Wacker}},
  \bibinfo{journal}{Adv. Solid State Phys.} \textbf{\bibinfo{volume}{41}},
  \bibinfo{pages}{199} (\bibinfo{year}{2001}).

\bibitem[{\citenamefont{Williams
  et~al.}(2005{\natexlab{a}})\citenamefont{Williams, Kumar, Hu, and
  Reno}}]{williams:copper}
\bibinfo{author}{\bibfnamefont{B.~S.} \bibnamefont{Williams}},
  \bibinfo{author}{\bibfnamefont{S.}~\bibnamefont{Kumar}},
  \bibinfo{author}{\bibfnamefont{Q.}~\bibnamefont{Hu}}, \bibnamefont{and}
  \bibinfo{author}{\bibfnamefont{J.~L.} \bibnamefont{Reno}},
  \bibinfo{journal}{Opt. Express} \textbf{\bibinfo{volume}{13}},
  \bibinfo{pages}{3331} (\bibinfo{year}{2005}{\natexlab{a}}).

\bibitem[{\citenamefont{Williams
  et~al.}(2003{\natexlab{b}})\citenamefont{Williams, Kumar, Callebaut, Hu, and
  Reno}}]{williams:metal}
\bibinfo{author}{\bibfnamefont{B.~S.} \bibnamefont{Williams}},
  \bibinfo{author}{\bibfnamefont{S.}~\bibnamefont{Kumar}},
  \bibinfo{author}{\bibfnamefont{H.}~\bibnamefont{Callebaut}},
  \bibinfo{author}{\bibfnamefont{Q.}~\bibnamefont{Hu}}, \bibnamefont{and}
  \bibinfo{author}{\bibfnamefont{J.~L.} \bibnamefont{Reno}},
  \bibinfo{journal}{Appl. Phys. Lett.} \textbf{\bibinfo{volume}{83}},
  \bibinfo{pages}{2124} (\bibinfo{year}{2003}{\natexlab{b}}).

\bibitem[{\citenamefont{Kohen et~al.}(2005)\citenamefont{Kohen, Williams, and
  Hu}}]{kohen:waveguide}
\bibinfo{author}{\bibfnamefont{S.}~\bibnamefont{Kohen}},
  \bibinfo{author}{\bibfnamefont{B.~S.} \bibnamefont{Williams}},
  \bibnamefont{and} \bibinfo{author}{\bibfnamefont{Q.}~\bibnamefont{Hu}},
  \bibinfo{journal}{J. Appl. Phys.} \textbf{\bibinfo{volume}{97}},
  \bibinfo{pages}{053106} (\bibinfo{year}{2005}).

\bibitem[{\citenamefont{Williams
  et~al.}(2005{\natexlab{b}})\citenamefont{Williams, Kumar, Hu, and
  Reno}}]{williams:corrdfb}
\bibinfo{author}{\bibfnamefont{B.~S.} \bibnamefont{Williams}},
  \bibinfo{author}{\bibfnamefont{S.}~\bibnamefont{Kumar}},
  \bibinfo{author}{\bibfnamefont{Q.}~\bibnamefont{Hu}}, \bibnamefont{and}
  \bibinfo{author}{\bibfnamefont{J.~L.} \bibnamefont{Reno}},
  \bibinfo{journal}{Opt. Lett.} \textbf{\bibinfo{volume}{30}},
  \bibinfo{pages}{2909} (\bibinfo{year}{2005}{\natexlab{b}}).

\bibitem[{\citenamefont{Maulini et~al.}(2006)\citenamefont{Maulini, Mohan,
  Giovannini, Faist, and Gini}}]{maulini:tunable}
\bibinfo{author}{\bibfnamefont{R.}~\bibnamefont{Maulini}},
  \bibinfo{author}{\bibfnamefont{A.}~\bibnamefont{Mohan}},
  \bibinfo{author}{\bibfnamefont{M.}~\bibnamefont{Giovannini}},
  \bibinfo{author}{\bibfnamefont{J.}~\bibnamefont{Faist}}, \bibnamefont{and}
  \bibinfo{author}{\bibfnamefont{E.}~\bibnamefont{Gini}},
  \bibinfo{journal}{Appl. Phys. Lett.} \textbf{\bibinfo{volume}{88}},
  \bibinfo{pages}{201113} (\bibinfo{year}{2006}).

\bibitem[{\citenamefont{Lee and Wacker}(2002)}]{lee:neg1}
\bibinfo{author}{\bibfnamefont{S.-C.} \bibnamefont{Lee}} \bibnamefont{and}
  \bibinfo{author}{\bibfnamefont{A.}~\bibnamefont{Wacker}},
  \bibinfo{journal}{Phys. Rev. B} \textbf{\bibinfo{volume}{66}},
  \bibinfo{pages}{245314} (\bibinfo{year}{2002}).

\bibitem[{\citenamefont{Wacker et~al.}(2009)\citenamefont{Wacker, Nelander, and
  Weber}}]{wacker:gain}
\bibinfo{author}{\bibfnamefont{A.}~\bibnamefont{Wacker}},
  \bibinfo{author}{\bibfnamefont{R.}~\bibnamefont{Nelander}}, \bibnamefont{and}
  \bibinfo{author}{\bibfnamefont{C.}~\bibnamefont{Weber}},
  \bibinfo{journal}{Proc. SPIE} \textbf{\bibinfo{volume}{7230}},
  \bibinfo{pages}{72301A} (\bibinfo{year}{2009}).

\bibitem[{\citenamefont{Khurgin et~al.}(2009)\citenamefont{Khurgin, Dikmelik,
  Liu, Hoffman, Escarra, Franz, and Gmachl}}]{khurgin:iroughness}
\bibinfo{author}{\bibfnamefont{J.~B.} \bibnamefont{Khurgin}},
  \bibinfo{author}{\bibfnamefont{Y.}~\bibnamefont{Dikmelik}},
  \bibinfo{author}{\bibfnamefont{P.~Q.} \bibnamefont{Liu}},
  \bibinfo{author}{\bibfnamefont{A.~J.} \bibnamefont{Hoffman}},
  \bibinfo{author}{\bibfnamefont{M.~D.} \bibnamefont{Escarra}},
  \bibinfo{author}{\bibfnamefont{K.~J.} \bibnamefont{Franz}}, \bibnamefont{and}
  \bibinfo{author}{\bibfnamefont{C.~F.} \bibnamefont{Gmachl}},
  \bibinfo{journal}{Appl. Phys. Lett.} \textbf{\bibinfo{volume}{94}},
  \bibinfo{pages}{091101} (\bibinfo{year}{2009}).

\bibitem[{\citenamefont{Scalari et~al.}(2003)\citenamefont{Scalari, Ajili,
  Faist, Beere, Linfield, Ritchie, and Davies}}]{scalari:btc}
\bibinfo{author}{\bibfnamefont{G.}~\bibnamefont{Scalari}},
  \bibinfo{author}{\bibfnamefont{L.}~\bibnamefont{Ajili}},
  \bibinfo{author}{\bibfnamefont{J.}~\bibnamefont{Faist}},
  \bibinfo{author}{\bibfnamefont{H.}~\bibnamefont{Beere}},
  \bibinfo{author}{\bibfnamefont{E.}~\bibnamefont{Linfield}},
  \bibinfo{author}{\bibfnamefont{D.}~\bibnamefont{Ritchie}}, \bibnamefont{and}
  \bibinfo{author}{\bibfnamefont{G.}~\bibnamefont{Davies}},
  \bibinfo{journal}{Appl. Phys. Lett.} \textbf{\bibinfo{volume}{82}},
  \bibinfo{pages}{3165} (\bibinfo{year}{2003}).

\bibitem[{\citenamefont{Kr\"{o}ll et~al.}(2007)\citenamefont{Kr\"{o}ll, Darmo,
  Dhillon, Marcadet, Calligaro, Sirtori, and Unterrainer}}]{kroll:nature}
\bibinfo{author}{\bibfnamefont{J.}~\bibnamefont{Kr\"{o}ll}},
  \bibinfo{author}{\bibfnamefont{J.}~\bibnamefont{Darmo}},
  \bibinfo{author}{\bibfnamefont{S.~S.} \bibnamefont{Dhillon}},
  \bibinfo{author}{\bibfnamefont{X.}~\bibnamefont{Marcadet}},
  \bibinfo{author}{\bibfnamefont{M.}~\bibnamefont{Calligaro}},
  \bibinfo{author}{\bibfnamefont{C.}~\bibnamefont{Sirtori}}, \bibnamefont{and}
  \bibinfo{author}{\bibfnamefont{K.}~\bibnamefont{Unterrainer}},
  \bibinfo{journal}{Nature} \textbf{\bibinfo{volume}{449}},
  \bibinfo{pages}{698} (\bibinfo{year}{2007}).

\bibitem[{\citenamefont{Jukam et~al.}(2008)\citenamefont{Jukam, Dhillon,
  Oustinov, Zhao, Hameau, Tignon, Barbieri, Vasanelli, Filloux, Sirtori
  et~al.}}]{jukam:gainbw}
\bibinfo{author}{\bibfnamefont{N.}~\bibnamefont{Jukam}},
  \bibinfo{author}{\bibfnamefont{S.~S.} \bibnamefont{Dhillon}},
  \bibinfo{author}{\bibfnamefont{D.}~\bibnamefont{Oustinov}},
  \bibinfo{author}{\bibfnamefont{Z.-Y.} \bibnamefont{Zhao}},
  \bibinfo{author}{\bibfnamefont{S.}~\bibnamefont{Hameau}},
  \bibinfo{author}{\bibfnamefont{J.}~\bibnamefont{Tignon}},
  \bibinfo{author}{\bibfnamefont{S.}~\bibnamefont{Barbieri}},
  \bibinfo{author}{\bibfnamefont{A.}~\bibnamefont{Vasanelli}},
  \bibinfo{author}{\bibfnamefont{P.}~\bibnamefont{Filloux}},
  \bibinfo{author}{\bibfnamefont{C.}~\bibnamefont{Sirtori}},
  \bibnamefont{et~al.}, \bibinfo{journal}{Appl. Phys. Lett.}
  \textbf{\bibinfo{volume}{93}}, \bibinfo{pages}{101115}
  (\bibinfo{year}{2008}).

\bibitem[{\citenamefont{Jukam et~al.}(2009)\citenamefont{Jukam, Dhillon,
  Oustinov, Mad\'{e}o, Tignon, Colombelli, Dean, Salih, Khanna, Linfield
  et~al.}}]{jukam:tdsphononqcl}
\bibinfo{author}{\bibfnamefont{N.}~\bibnamefont{Jukam}},
  \bibinfo{author}{\bibfnamefont{S.~S.} \bibnamefont{Dhillon}},
  \bibinfo{author}{\bibfnamefont{D.}~\bibnamefont{Oustinov}},
  \bibinfo{author}{\bibfnamefont{J.}~\bibnamefont{Mad\'{e}o}},
  \bibinfo{author}{\bibfnamefont{J.}~\bibnamefont{Tignon}},
  \bibinfo{author}{\bibfnamefont{R.}~\bibnamefont{Colombelli}},
  \bibinfo{author}{\bibfnamefont{P.}~\bibnamefont{Dean}},
  \bibinfo{author}{\bibfnamefont{M.}~\bibnamefont{Salih}},
  \bibinfo{author}{\bibfnamefont{S.~P.} \bibnamefont{Khanna}},
  \bibinfo{author}{\bibfnamefont{E.~H.} \bibnamefont{Linfield}},
  \bibnamefont{et~al.}, \bibinfo{journal}{Appl. Phys. Lett.}
  \textbf{\bibinfo{volume}{94}}, \bibinfo{pages}{251108}
  (\bibinfo{year}{2009}).

\bibitem[{\citenamefont{Kumar et~al.}(2008)\citenamefont{Kumar, Lee, Qin,
  Williams, Hu, and Reno}}]{kumar:ptw08}
\bibinfo{author}{\bibfnamefont{S.}~\bibnamefont{Kumar}},
  \bibinfo{author}{\bibfnamefont{A.~W.~M.} \bibnamefont{Lee}},
  \bibinfo{author}{\bibfnamefont{Q.}~\bibnamefont{Qin}},
  \bibinfo{author}{\bibfnamefont{B.~S.} \bibnamefont{Williams}},
  \bibinfo{author}{\bibfnamefont{Q.}~\bibnamefont{Hu}}, \bibnamefont{and}
  \bibinfo{author}{\bibfnamefont{J.~L.} \bibnamefont{Reno}},
  \bibinfo{journal}{Proc. SPIE} \textbf{\bibinfo{volume}{6909}},
  \bibinfo{pages}{69090I} (\bibinfo{year}{2008}).

\bibitem[{\citenamefont{Dunbar et~al.}(2007)\citenamefont{Dunbar, Houdr\'{e},
  Scalari, Sirigu, Giovannini, and Faist}}]{dunbar:microcavity}
\bibinfo{author}{\bibfnamefont{L.~A.} \bibnamefont{Dunbar}},
  \bibinfo{author}{\bibfnamefont{R.}~\bibnamefont{Houdr\'{e}}},
  \bibinfo{author}{\bibfnamefont{G.}~\bibnamefont{Scalari}},
  \bibinfo{author}{\bibfnamefont{L.}~\bibnamefont{Sirigu}},
  \bibinfo{author}{\bibfnamefont{M.}~\bibnamefont{Giovannini}},
  \bibnamefont{and} \bibinfo{author}{\bibfnamefont{J.}~\bibnamefont{Faist}},
  \bibinfo{journal}{Appl. Phys. Lett.} \textbf{\bibinfo{volume}{90}},
  \bibinfo{pages}{141114} (\bibinfo{year}{2007}).

\end{thebibliography}

\end{document}